%% file: main.tex
\documentclass[sigplan,nonacm]{acmart}
\input{setup}
\author{Martin Kong}
\affiliation{%
  \institution{The University of Oklahoma}
  \country{Norman, OK, USA}
}
\email{mkong@ou.edu}
\renewcommand{\shortauthors}{Martin Kong}


\begin{document}

\thispagestyle{empty}

\title{\mytitle}

\setcopyright{none}
\copyrightyear{}
\acmYear{}
\acmDOI{}
\acmConference[]{}{}{}

\acmPrice{}
\acmISBN{}

\date{}
\input{abstract}

\maketitle

\section{Introduction}
\label{sec:intro}
\input{intro}

\section{\label{sec:background}Background in Quantum Computing}

\input{background}

\section{\label{sec:polyhedral}The Polyhedral Model in Quantum Computing}

\input{abstractions}

\subsection{\axl: A simple Domain-Specific Language for Parameterized Affine Quantum Circuits}

\input{lang}
\subsection{Transformations}

\input{transfos}
\section{Experimental Evaluation}
\label{sec:results}
\input{results}
\section{Related Work}
\label{sec:related}

\input{related}

\section{Conclusion and Future Directions}
\label{sec:conclusion}
\input{conclusion}

\bibliographystyle{ACM-Reference-Format}
\balance
\bibliography{allrefs}

\begin{appendices}
\section{}
\label{sec:appendix}

\input{appendix}

\end{appendices}

\end{document}

%% file: setup.tex
\usepackage[normalem]{ulem}
\usepackage{booktabs}   
\usepackage[shortcuts]{extdash}
\usepackage{afterpage}
\usepackage{setspace}
\usepackage[multiple]{footmisc}
\usepackage{url}
\usepackage[title]{appendix}
\usepackage{alltt}
\usepackage{xspace}
\usepackage{epsfig}
\usepackage{graphics}
\usepackage{bbold}
\usepackage{empheq}
\usepackage{caption}
\usepackage{colortbl,hhline}
\usepackage{multirow}
\usepackage{tikz}

\usepackage{amsmath,amsfonts,amsbsy}
\usepackage{fancyvrb,keyval,ifthen}
\usepackage{graphicx}

\usepackage{booktabs}
\usepackage{listings}
\usepackage{enumitem}
\setlist{nolistsep}
\usepackage{titlesec}
\usepackage{balance}
\usepackage{array}
\usepackage{subfig}

\usepackage[linesnumbered,ruled,vlined]{algorithm2e}

\newcommand{\axl}[0]{{\bf{\texttt{AXL}}}}

\newcommand{\bcheung}[0]{\texttt{cheung}}
\newcommand{\bcuccaro}[0]{\texttt{cuccaro-adder-6bit}}
\newcommand{\bpipelined}[0]{\texttt{pipelined-swap}}
\newcommand{\bcnt}[0]{\texttt{cnt3-5\_179}}

\newcommand{\baddermau}[0]{\texttt{adder-maj-uma}}
\newcommand{\bsum}[0]{\texttt{sum5}}
\newcommand{\binit}[0]{\texttt{init-G$_5$}}
\newcommand{\bparity}[0]{\texttt{parity\_247}}
\newcommand{\brd}[0]{\texttt{rd84\_142}}
\newcommand{\topogrid}[0]{\texttt{grid}}
\newcommand{\topomring}[0]{\texttt{multi-ring}}
\newcommand{\topotiled}[0]{\texttt{tiled}}
\newcommand{\trbase}[0]{\textit{base}}
\newcommand{\trfeautrier}[0]{\textit{feautrier}}
\newcommand{\trplutomin}[0]{\textit{plutomin}}
\newcommand{\trplutomax}[0]{\textit{plutomax}}

\newcommand{\mytitle}[0]{Exploring the Impact of Affine Loop Transformations in Qubit Allocation}

%% file: abstract.tex
\begin{abstract}
Most quantum compiler transformations and qubit allocation techniques to date
are either peep-hole focused or rely on sliding windows that depend on a
number of external parameters.  Thus, global optimization criteria are still
lacking.
In this paper we explore the synergies and impact of affine loop
transformations in the context of qubit allocation and mapping.  With this goal
in mind, we have implemented a domain specific language and source-to-source
compiler for quantum circuits that can be directly described with affine
relations. We conduct an extensive evaluation spanning 8 quantum circuits taken
from the literature, 3 distinct coupling graphs, 4 affine transformations
(including the Pluto dependence distance minimization and Feautrier's minimum
latency algorithms), and 4 qubit allocators.  Our results demonstrate that
affine transformations using global optimization criteria can cooperate effectively in
several scenarios with quantum qubit mapping algorithms to reduce the
circuit depth, size and allocation time. 

\end{abstract}

%% file: intro.tex
The field of Quantum Computing (QC) has made tremendous advances in the last 
two decades at the hardware (e.g. ion trap and superconducting QC), algorithmic (QFT 
\cite{barenco.physreva.1996,chiaverini.science.2005,cleve.fcs.2000,hales.fcs.2000}, 
Grover's Search \cite{grover.toc.1996,kwiat.jomp.2000,boyer.pophys.1998}, 
Shor's algorithm \cite{shor.siam.1999,shor.fcs.1994,martin.nature.2012,lanyon.physreva.2007,beauregard.arxiv.2002}), and software levels \cite{ibm-quantum-experience.nature.2017,openqasm.arxiv.2017,openfermion.arxiv.2017}.
Known quantum algorithms already provide us with a glimpse of their expected
exceptional complexity. Thus, it is imperative for a programming language to be
a vehicle for algorithmic specification rather than an obstacle in the path to
progress. 
To address and bridge the semantic gap between algorithm specification and
quantum architectures, several languages, compilers and frameworks have been
proposed. Examples of these are ProjectQ \cite{projectq.2018}, Scaffold and the
ScaffCC compiler \cite{scaffcc.parco.2015}, Quipper \cite{quipper.pldi.13},
Microsoft's Q\# DSL \cite{qsharp.rwdslw.2018} and SIMD approaches such as
\cite{martonosi.asplos.2015}, or approaches focused on safe uncomputation such as SILQ \cite{silq.pldi.2020}.

Ultimately, the high-level programming language produces a stream of quantum
assembly operations \cite{openqasm.arxiv.2017,qasm.2017}, at which point \emph{Qubit
Allocation}, a technique akin to classic register allocation \cite{chaitin.cc.1982,poletto.toplas.1999}, is applied to
find a space-time mapping of the quantum gate operations in the program to the
quantum device. Qubit allocation techniques typically decompose the input
program into (network) layers, and generally suffer from limitations such as
approximating the global solution from local optima \cite{openqiskit.2018}, 
use relative small sliding windows \cite{sabre.asplos.2019}, 
leverage random initial mappings \cite{siraichi.cgo.2018,openqiskit.2018}, 
or incur in high time and space
complexity due to exponentially large search spaces \cite{jku.date.2018,astar-search.1968}.

The main goal of this work is to explore and understand potential synergistic
interactions between affine loop transformations and qubit allocation techniques
in order to find scenarios where the power of a global optimization criteria
can effectively improve the quality of the qubit allocation.
Our evaluation shows that even state-of-the-art allocators such as {\bf sabre} \cite{sabre.asplos.2019}
can improve by as much as 34\% with classical polyhedral loop transformations,
while other techniques less computationally demanding (i.e. {\bf wpm} \cite{siraichi.cgo.2018}) 
can improve by up to 60\%\footnote{{\bf sabre}'s average improvement over {\bf jku} for large circuits is 14\%, reported in \cite{sabre.asplos.2019}}.



In summary, this paper makes the following contributions:
i) We conduct an extensive study to understand the interactions
of affine loop transformations with qubit allocation techniques.
Our evaluation encompasses 8 quantum circuits, 4 allocators, 3 topologies
and 3 affine transformations in addition to the pass through code generation mode.
ii) Inspired in the Omega calculator \cite{omega.tr.1996} and ISCC's \cite{iscc.impact.2011}
notation, we introduce a simple domain specific language based on polyhedral 
abstractions to enable the description, manipulation and composition
of quantum circuits.
iii) We discuss how the polyhedral model can be used as an efficient
intermediate representation for the optimization of affine quantum circuits.
In particular, we highlight how we can use it to represent quantum networks.


The rest of this paper is organized as follows. 
Sec.\ref{sec:background} recaps the necessary quantum terminology and background. 
In particular, we briefly recap several quantum allocators recently introduced, and
which we use in our evaluation.
Sec.\ref{sec:polyhedral} revises the polyhedral background while introducing
a simple domain-specific language for affine quantum circuits.
Sec.\ref{sec:results} discusses our extensive evaluation, general results,
individual analyses and scalability tests.
We conclude this paper with the related work (Sec.\ref{sec:related}) and the final remarks 
in Sec.\ref{sec:conclusion}.

%% file: background.tex
\paragraph{Qubits}
Quantum bits are the basic unit of information of quantum programs,
and are the analogous of classical bits. However, unlike their classical
counterpart, which can only take the values in the set \{0,1\}, qubit
values take the form of linear combinations of two basis states ($|0\rangle$ and $|1\rangle$).

\paragraph{Gates and Measurement}
Are the elementary operations applied to qubits. Their role is to
evolve the state of qubits. 
More generally, quantum gates can be seen as unitary matrix
operations applied to vectors representing quantum states.
Current quantum technologies utilize
gate operations with 1 and up to 3 qubits. An example of a single
qubit operation is the NOT (X) gate, which negates the state of a single
qubit. An example of a two-qubit gate is the CNOT (Controlled-NOT or CX) operation, 
which reverses the state of the second qubit operand when the first one is 1.
An example of a 3-qubit gate is the CCNOT (Controlled-Controlled-NOT), which
utilizes two control and one target qubits. In terms of classical computing
the control-qubits are {\em read-only} while the target qubit is effectively {\em updated}.
Quantum gates are akin to assembly code in classical computing.

\paragraph{Coupling Graphs}
A quantum processor can be conceptually conceived as a graph/network where the vertices
are the qubits, and the edges the communication links between them.
Computational steps (gates) performed in the network are synchronized in time.
Qubits serve as inputs and outputs to each quantum operation.
A current limitation of quantum computing technology, is that multi-qubit
gates can only be applied to qubits directly connected by a link \cite{xiang.romp.2013,ladd.nature.2010}.

\paragraph{Quantum Circuits}
Quantum programs can be graphically represented in the form of circuits
\cite{vedral.1996}.  We show an example of a simple circuit obtained from the
Revlib online repository \cite{revlib-web} in Fig.\ref{fig:parity}.  The x-axis
signifies time, while the y-axis are the qubits available in the quantum
processor.  It uses 6 qubits, 5 NOT gates (left-most), and 5 CNOT gates. Each
NOT gate is synchronized with the control qubit of a CNOT gate.  Operations on
the same qubit lane, from left-to-right execute one-after another, and embody
classic data-flow input/output dependences. 

\begin{minipage}{\linewidth}
\begin{minipage}{0.28\linewidth}
\includegraphics[width=\linewidth]{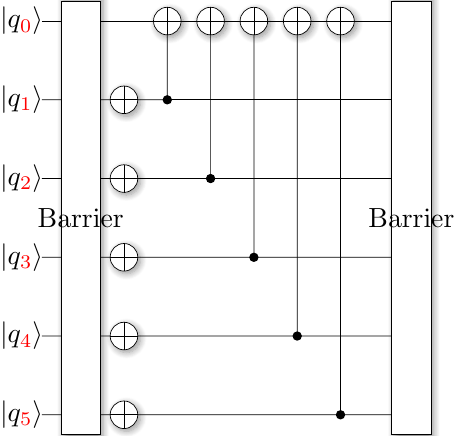}
\captionof{figure}{\label{fig:parity}parity\_247 circuit \cite{revlib-web}}
\end{minipage}
\begin{minipage}{0.66\linewidth}
The actual semantics of the specific
gate determine if the state of a qubit $q_i$ is read, written, or both.
The {\em depth} of a quantum circuit is the maximum number of gates 
scheduled on any single qubit lane, whereas the total number of gates is the {\em circuit size}. 
The quantum circuit ultimately defines the
unitary evolution of the input (initial) state into the final/output state.
\end{minipage}
\end{minipage}


\paragraph{Qubit Allocation}
Is the space-time mapping (assignment) of quantum operations to qubits in the
coupling graph, and  is very similar in spirit to the classical register allocation problem.
Recently several qubit allocators have been proposed. We briefly summarize
a few of the most relevant techniques.
The {\bf ibm-mapper} \cite{openqiskit.2018} (available in Qiskit) 
divides the input into a sequence of layers using disjoint sets of qubits;
qubits within layers are mapped by minimizing the sum of squared distances among vertices, and
potentially inserting SWAP operations. In addition, each distance term is scaled by
a factor $1 + r$, where $r$ is a random number between 0 and 1. The algorithm defaults
to one gate per layer if a valid mapping is not found.
{\bf wpm} \cite{siraichi.cgo.2018} is a heuristic that finds first an initial
allocation maximizing the number of control dependences, followed by a 
second pass which completes any remaining ordering constraint, potentially
inserting SWAP operations. Siraichi et al. \cite{siraichi.cgo.2018} also 
proposed an exact solution to the qubit allocation problem as a dynamic program
with state memoization, and showing that it leads to a $O(|Q|!^2~.~|Q|~.~|D|)$,
where $Q$ is the set of (physical) qubits, and $D$ the list of dependences. The
heuristic proposed ({\bf wpm}) was shown to achieve 
$O(|Q|~.~lg(|Q|) + |E| + |D| + |D|~.~(|Q|+|E|))$.
{\bf jku} \cite{jku.date.2018} uses the circuit size as the main optimizing metric,
decomposing the input network into layers, and attempting to minimize
the number of gates within each layer. To avoid falling into local optima,
{\bf jku} uses an $A*$ search algorithm \cite{astar-search.1968}, a family of 
graph traversal algorithms known to incur in $O(b^d)$ space complexity,
where $b$ is the branching factor in a tree and $d$ its depth. The heuristic used to
drive the search attempts to convert the mapping of each layer into the subsequent
one by inserting SWAP operations.
{\bf sabre} \cite{sabre.asplos.2019} is an efficient algorithm with $O(N^{2.5}g)$ time complexity
that attempts to minimize both the circuit depth
and size, usually exploring a trade-off between them. As previous techniques, it
divides the input program into layers, but performs two additional passes (a back-traversal
and a second forward traversal) to improve the initial (random) mapping.

\paragraph{(De)Coherence}
One of the main challenges facing QC is the decoherence problem, where the state
of a qubit decays over time. Each quantum gate has a specific decoherence time,
which for state-of-the-art quantum machines using superconducting technology
is approx.~100 $\mu$s \cite{ibm-quantum-experience.nature.2017}. In addition, 
gate operations can also introduce errors at rates varying between $O(10^{-3})$ for single-qubit gates and
$O(10^{-2})$ for two-qubit gates. The execution time of a quantum circuit results
from the aggregated time needed to run all the gates along the circuit's depth.
Thus, minimizing this metric is one of the main optimizing criteria.
Likewise, minimizing the total number of operations in circuits also equates to reducing
the compounded error.



%% file: abstractions.tex
In this section we quickly revise the 4 polyhedral abstractions
in the context of (affine) quantum circuits. We then describe
a simple domain-specific language heavily inspired in the
Omega Calculator \cite{omega.tr.1996} and ISCC \cite{iscc.impact.2011}
notation.  However, unlike its predecessors, one of its main goals
is to facilitate and capture the identity schedule of the quantum circuit.
We explain how the circuit structure and dependences
are mapped to the polyhedral abstractions, and quickly recap
well known affine loop transformations previously proposed.

\subsection{Polyhedral Abstractions}

Polyhedral compilers focus on fragments of programs that exhibit
static control parts (SCoPs)\cite{feautrier.ijpp.1991,feautrier.ijpp.1992a,feautrier.ijpp.1992b}. 
From these, four abstractions are extracted: iteration domains, access functions,
dependence polyhedra and scattering/scheduling functions
\cite{griebl.ijpp.2000}.

\input{gates}

\textbf{Iteration Domains:}
Each syntactic statement $S$ is associated
to the set of points $\mathcal{D}^S$ in $\mathbb{Z}^{+}$ comprised by the
dynamic instances of the statement. In the context of QC, an iteration
domain can group several operations of the same type or of different ones,
but that behave in the same fashion. Consider for instance
the \bparity~ circuit shown in Fig.\ref{fig:parity}, which can be represented
with two iteration domains, $\mathcal{D}^{S1} = \{ [i] : 0 \leq i < N \}$ and
$\mathcal{D}^{S2} = \{ [i] : 0 \leq i < N \}$, one for the NOT operations,
and another for the CNOTs; where $N$ is an unknown but fixed value
that parameterizes the circuit.
We also note that several techniques have been developed to model and extend
the applicability of the polyhedral model to different forms of irregular computations
\cite{venkat.cgo.2014,venkat.pldi.2015,augustine.pldi.2019}.
The remaining three abstractions are
essentially functions applied to the iteration domains.

\textbf{Program schedules:}
The execution order of statement instances is defined in the polyhedral
model with the program schedule, a transformation matrix or an affine map
that assigns to each statement instance an execution date.
Schedules can be seen as multi-dimensional time-stamps 
\[\Theta^S(\vec{x}) = \left<\theta^S_1, \theta^S_2, \ldots,\theta^S_d\right>,~~
\vec{x} \in \mathcal{D}^S,\]
where $\theta^S_i$ is a one-dimensional affine function, and
$d$ is the number of dimensions in $\Theta^S$.
Schedules can be lexicographically compared, and are used in the
polyhedral scanning process \cite{bastoul.pact.2004} to generate
the loop structure that will visit each statement instance in the 
order established by the schedule $\Theta^S$. Continuing with our ongoing example,
the execution order of \bparity~ circuit can be represented 
with the schedules $\Theta^{S1}(i) = \left<0,i\right>,~\Theta^{S2}(i) = \left<1,i\right>$.
We note, however, that for this circuit loop fusion can be legally applied.
In this case, another potential schedule could be
$\Theta^{S1}(i) = \left<0,i,0\right>,~\Theta^{S2}(i) = \left<0,i,1\right>$.


\textbf{Access Relations:}
Are an abstraction that permit to model memory accesses. 
Access relations map points in an iteration domain to a data-space.
The motivation is two-fold.
First, to
later be able to identify program statements accessing the same memory location.
Second, to update the array subscript functions post-transformation.
The construction of access relations in quantum computing differs in two
fundamental ways from its classical counterpart.  First, in classical computing
the usage of multi-dimensional arrays is the norm, whereas the current practice
in quantum computing is to operate on a single, large, one-dimensional array
that represents a quantum register. The second difference involves what
constitutes a \textit{read} and \textit{write} access. In effect, in
classical computing, where polyhedral compilation has been predominantly
applied to imperative programming languages such as C/C++ and Fortran,
there is, practically  always, a single write reference and zero or more read
references.  This is not the case in quantum computing, where the program's
state evolves as specific entries of the register are input to gate
operations. Moreover, the type of gate determines if a specific register
entry is \textit{read} or \textit{updated}. We make this distinction for
two reasons: First, some gates take {\textit control argument}, which
do not modify the contents of a register entry. Second, every gate operation
performing a \textit{write} also reads the input register entry, i.e.
the same entry is both read and written. Resuming our example,
if we assume that the top qubit has index 0, increasing downwards, then
statement S2 would have three access relations, a read and write relation
$\{ [i] \rightarrow [0] \}$
for the target qubit, and a read-only relation $\{ [i] \rightarrow [i+1] \}$ for the control qubit.

\textbf{Dependence Polyhedra}
embody the semantic orderings of the program.
Every program dependence in a SCoP is represented by one or more
dependence polyhedra $\mathcal{D}^{R \rightarrow S}$. 
These polyhedra define
the ordering among points $\vec{x}^{R}$ and $\vec{y}^{S}$
from the iteration domains $\mathcal{D}^{R}$
and $\mathcal{D}^{S}$, respectively.
This critical pass is necessary to perform aggressive loop optimizations,
and compute reordering transformations via one or more {\it integer linear problems} (ILPs),
which must preserve the legality of the transformations applied.
Essentially every scheduling technique 
\cite{feautrier.rairo.1988,vasilache.ics.2005,vasilache.07.phd,girbal.ijpp.2006,acharya.pldi.2018}
embeds the program semantic constraints in the form of dependence polyhedra
(possibly linearized by the application of the Farkas Lemma) into one or more ILP
systems. These polyhedra are usually extracted with ``classical'' data-flow dependence analysis
\cite{feautrier.ijpp.1991}.

\textbf{(Parameterized) Affine Quantum Circuits (PAQCs)}
We define PAQCs as a subclass of quantum programs that can be expressed
with affine relations, or a union of them. We have two requirements:
First, the instances of gate operations to be groupable by an affine expression;
Second, the indices of qubits being accessed to be representable via affine functions.

%% file: gates.tex
\begin{table*}[t]
\centering
\begin{footnotesize}
\caption{\label{tab:gates} Quantum Gates}
\begin{tabular}{|c|c|c|c|c|}
\hline
Gate & Description & No.Inputs & No.Outputs & Integer Map Representation\\
\hline

$X(rw_1)$ & Pauli-X (NOT) gate & 1 & 1  & S$[\vec{x}]$ $ \rightarrow $ q[$rw_1(\vec{x})$] \\

$Y(rw_1)$ & Pauli-Y gate & 1 & 1 & S$[\vec{x}]$ $ \rightarrow $ q[$rw_1(\vec{x})$] \\

$Z(rw_1)$ & Pauli-Z gate& 1 & 1 & S$[\vec{x}]$ $ \rightarrow $ q[$rw_1(\vec{x})$] \\

$H(rw_1)$ & Hadamard (H) gate & 1 & 1 & S$[\vec{x}]$ $ \rightarrow $ q[$rw_1(\vec{x})$] \\

$Measure(rw_1,w_2)$ & Measure qubit to classical bit  & 1 & 2 & 
  S$[\vec{x}]$ $ \rightarrow $ q[$rw_1(\vec{x})$],~~  S$[\vec{x}]$ $ \rightarrow $ c[$w_2(\vec{x})$] \\

$CNOT(r_1,rw_2)$ & Controlled-NOT (CX) & 2 & 1 & 
  S$[\vec{x}]$ $ \rightarrow $ q[$r_1(\vec{x})$],~~  S$[\vec{x}]$ $ \rightarrow $ q[$rw_2(\vec{x})$] \\

$CY(r_1,rw_2)$ & Controlled-Y & 2 & 1 & 
  S$[\vec{x}]$ $ \rightarrow $ q[$r_1(\vec{x})$],~~ S$[\vec{x}]$ $ \rightarrow $ q[$rw_2(\vec{x})$] \\

$CZ(r_1,rw_2)$ & Controlled-Z & 2 & 1 & 
  S$[\vec{x}]$ $ \rightarrow $ q[$r_1(\vec{x})$],~~  S$[\vec{x}]$ $ \rightarrow $ q[$rw_2(\vec{x})$] \\

$Swap(rw_1,rw_2)$ & Exchange state & 2 & 2 & 
  S$[\vec{x}]$ $ \rightarrow $ q[$rw_1(\vec{x})$],~~  S$[\vec{x}]$ $ \rightarrow $ q[$rw_2(\vec{x})$] \\

$Toffoli(r_1,r_2,rw_3)$ & Controlled-Controlled-Not & 2 & 1 & 
  S$[\vec{x}]$ $ \rightarrow $ q[$r_1(\vec{x})$],~~
  S$[\vec{x}]$ $ \rightarrow $ q[$r_2(\vec{x})$],~~
  S$[\vec{x}]$ $ \rightarrow $ q[$rw_3(\vec{x})$] \\

\hline
\end{tabular}
\end{footnotesize}
\end{table*}

%% file: lang.tex
Enter \axl, a declarative
language with operations that enable the
specification, manipulation and composition of quantum circuits.  
\axl's syntax is simple and straight-forward.
\axl~ provides 4 datatypes: {\it gate}, {\it circ} 
(constant circuits), {\it statement} circuits (our link to
the polyhedral abstractions) and program parameters. The former
three define a hierarchy of types: $gates \subset circ \subset statement$.
A gate is any of the operations shown in Table \ref{tab:gates},
and represents a single quantum assembly instance (or a point
in $\mathbb{Z}^+$).
The {\it circ} type essentially allows for the composition
of gates. For example, the NOT gate operating on qubit 1 can be written
as NOT(1),  and a CNOT gate controlled by qubit 0 and targeting qubit 1
can be expressed as CNOT(0,1). These operations can then be synchronized
(in time) via the time composition operator $(+)$. Internally, \axl~
will manipulate the program schedule to properly represent this ordering
constraint. Climbing the hierarchy we have a {\it circuit statement}, 
an enriched {\it circ}
with polyhedral abstractions (iteration domains, access relations,
gate relations, schedules). Due to space constraints, in this paper
we will focus our attention on the {\it statement} type.
%

Figure \ref{lst:axl-parity} shows the code producing two 
implementations of the \bparity~ circuit in
\axl.
The first one uses a single statement (S1)
to model the entire circuit, embedding both gates in its body.
The (+) time-composition operator synchronizes the NOT and CNOT
operations for every $t \in \mathcal{D}^{S1}$. In contrast,
the second implementation puts each gate in its own statement (S2 and S3).
Lines 6-7 performs code generation from each SCoP while providing
non-default parameters and choosing a pre-determined transformation
of choice. The second implementation imposes the synchronization
of both statements also by the time-composition operator (+),
which internally introduces a leading scalar dimension to their respective
program schedule.

\begin{figure}[h]
\lstset{language=C,xleftmargin=3em,frame=single,framexleftmargin=0em,
basicstyle=\footnotesize\ttfamily,
  keywordstyle=\color{blue}, frame=single,numbers=left,
}%
\begin{lstlisting}
param M;
statement S1, S2, S3;
S1 := {t:1<=t<=M (%) #NOT(t) (+) #CNOT(t,0)};
S2 := {t:1<=t<=M (%) #NOT(t) };
S3 := {t:1<=t<=M (%) #CNOT(t,0)};
codegen {S1}  with {M=8} apply {plutomax};
codegen {S2(+)S3} with {M=8} apply {plutomin};
\end{lstlisting}
\caption{\label{lst:axl-parity}\axl~ implementations of circuit \bparity}
\end{figure}

\textbf{Parameterized Affine Quantum Circuits (PAQC) }
A unique capability of \axl~ is to represent {\it affine quantum
circuits} in a parameterized fashion, that is, circuits can be 
modeled with symbolic values representing arbitrary, unknown,
constant values. PAQCs are built by defining an iteration
domain for the circuit and by attaching a circuit body to it.
The circuit body can be: i) a basic gate; ii) a fixed, possibly composite
expression of type {\bf circ}; or iii) a parameterized, variable {\bf circ}
expression. The last of these body types, leverages the full power of the
polyhedral model to represent individual instances of the computation.
PAQCs are implemented in \axl~ via the {\bf statement} type,
making parameterized and variable sized quantum circuits first class citizens
in our language. 

We show in Table \ref{tab:gates} a subset of the gates supported by \axl.
The arguments to each gate represent indices in an implicit one-dimensional space;
depending on the type of the gate, some arguments are either {\em read-only}
(i.e. any $r_i$ argument), {\em read-write} (arguments $rw_j$) or {\em write-only}
(arguments $w_k$). These semantics allow us to determine the number of input
and output relations, which we show in the third and fourth columns.
The last column shows the extracted access relations for each of the accessed
register entries, modeled as a single affine relation mapping points of the iteration
domain to the data-space of the quantum register (register {\bf q}) or of the
classical register (register {\bf c}). We follow the same naming convention
to indicate when an access function is of type read, read-write or write.
We note here that, although we list the main quantum gate operations,
these can be quickly added into the \axl~ language under the category
defined by the number of qubits, and the access type to each of its
arguments.

\textbf{Gate Call Relations}
In classical (scientific) computing, where multi-dimensional arrays are
pervasive in imperative programming languages, each array reference is either
read or written.  This naturally follows the semantics of languages with
explicit assignment operations; only the reference on the left-hand-side is
written, while any array or scalar variable on the right-hand-side is read-only
(except when passed to some function with side-effects). In our DSL, quantum
programs operate in an implicit 1-dimensional data-space. While there 
is no inherent limitation in \axl~to represent registers in a multi-dimensional space,
most quantum technologies have assumed this so far. 
Thus, we add a new abstraction representing {\em quantum gate calls}. These
are very similar in spirit to classical access relations that map points
in iteration domains to the various data spaces of the program. While quantum
gates are effectively functions that could change the state of a register,
they do come equipped with domain semantics that must be correctly mapped.
More precisely, each quantum gate operation has a set of {\em read} and {\em write}
index sets that establish the register entries accessed in each of these modes.
Later, during the polyhedral scanning process (phase-1 of code generation)
these relations are used to produce the exact quantum gate operation specified
by the end-user.

\textbf{Program Assembly}
We define {\bf Program Assembly} as the process of constructing the final 
circuit schedule establishing the complete ordering among statement circuit
instances in a single SCoP. It takes place at the start of the transformation and code generation
process (see \texttt{codegen} clause in Fig.\ref{lst:axl-parity}). 
Building the final schedule happens in two phases. First, when defining
circuit statements, a local identity schedule is created.
For instance, in the same example, the local schedule 
$\Theta = \left<i\right>$ is created for statements S1-S3 (Lines 3-5 in the same figure).
\axl~ will then split S1 into sub-statements S1$_1$ and S1$_2$, effectively making each gate
its own statement.
The usage of the (+) operator for S1 will introduce a suffix scalar dimension,
producing the schedules $\Theta^{S1_1} = \left<i,0\right>$ and
$\Theta^{S1_2} = \left<i,1\right>$.
The second phase of program assembly synchronizes statement circuits in a similar fashion,
but with the distinction of inserting scalar dimensions as prefixes.
For example, the local schedules of S2 and S3 will be modified to
$\Theta^{S2} = \left<0,i\right>$ and
$\Theta^{S3} = \left<1,i\right>$.
Along the process, multiple scalar dimensions might be added to faithfully
represent the ordering among various circuit patterns, which can be controlled
by proper parenthesization.

%% file: transfos.tex
We next recap two well known formulations that have been
extensively used at the heart of several scheduling algorithms,
the Feautrier minimun latency schedules \cite{feautrier.ijpp.1992a,feautrier.ijpp.1992b}, and the minimization of 
the maximal dependence distance used in the Pluto compiler \cite{uday.pldi.2008}.
In particular, Pluto has been extremely successfully in generalizing
the tileability of imperfectly nested loops and using the aforementioned
cost function to produce high-quality transformations. Similarly, the Feautrier
algorithm has been used in \cite{kong.pldi.2013} and is still the fall-back
scheduling strategy in ISL \cite{isl.2010}.

\begin{figure}[h]
\lstset{language=C,xleftmargin=3em,frame=single,framexleftmargin=0em,
basicstyle=\footnotesize\ttfamily,
  keywordstyle=\color{blue}, frame=single,numbers=left,
}%
\begin{lstlisting}
// Plutomin (min.fusion) heuristic: 
// Loop fission
for (int c1 = 1; c1 <= 8; c1 += 1) {
  X[c1];
}
for (int c1 = 1; c1 <= 8; c1 += 1) {
  CX[c1][0];
}
// **********************************
// Plutomax (max.fusion) heuristic: 
// Loop fusion
for (int c0 = 1; c0 <= 8; c0 += 1) {
  X[c0];
  CX[c0][0];
}
\end{lstlisting}
\caption{\label{lst:parity-loop}Generated loop structure for circuit \bparity~ using Pluto's
Minfuse and Maxfuse loop fusion heuristics}
\end{figure}

\paragraph{The Pluto Algorithm}
The Pluto Tiling Hyper-plane algorithm \cite{uday.pldi.2008}, was introduced
as a general greedy algorithm to find affine transformations to make a program tileable.
At its core, it uses a cost function that bounds and minimizes 
the distance among statements in dependence relations. As we don't
apply loop tiling to quantum programs, we only recap Pluto's main
cost function formulation below:
\begin{eqnarray*}
\delta_{e}(\vec{x}) = \phi_{S_i}(\vec{x}) - \phi_{S_j}(h_{e}(\vec{x})),~~
\vec{x} \in P_{e}
\\
\phi_{S_i}(\vec{x}) - \phi_{S_j}(h_{e}(\vec{x})) \leq v(\vec{p}),~~
\vec{x} \in P_{e}, ~~ \forall e \in E
\\
v(\vec{p}) = u . \vec{p} + w
\\
v(\vec{p}) - \delta_{e}(\vec{x}) \geq 0, ~~
\vec{x} \in P_{e}, ~~ \forall e \in E
\\
minimize_{\prec} \{ u=(u_1,u_2,u_3,\ldots,u_n),\ldots \}
\end{eqnarray*}
The above equations essentially perform the following: 
i) define the constraints to satisfy a dependence edge $e$ of the dependence
set $E$, for every point $\vec{x}$ in the dependence polyhedron $P_{e}$;
ii) introduce an affine function $v$ on the vector of program parameters, $\vec{p}$;
iii) bound the distance between the target and source of the dependence
via a function $v(\vec{p})$;
iv) minimize the \texttt{u} coefficients that bound the dependence distance
(NOTE: we omit here the application of the Farkas Lemma).

The rest of Pluto's algorithm proceeds level-by-level, from the outermost to
the innermost, finding one-dimensional affine transforms for all statements in
the program. By default, {\em splitters} (scalar dimensions) are only introduced
by the algorithm when no solutions are found for a new hyperplane. However,
doing this whenever is legal, produces loop structures maximally distributed.
We show in Fig.\ref{lst:parity-loop} the result of applying both heuristics,
labelled as {\em Plutomin} and {\em Plutomax}, to the \axl~ implementation
of the \bparity~ circuit.


\input{table-results-summary}

\paragraph{The Feautrier Scheduling Algorithm}
Feautrier's seminal papers \cite{feautrier.ijpp.1992a,feautrier.ijpp.1992b}
on one-dimensional and multi dimensional affine schedules have been
a common test for several applications and domains. The main property
of Feautrier's scheduling approach is to greedily satisfy as many dependences
as early as possible, going from the outermost (linear) dimension to the innermost.
This greedy approach produces schedules with the minimum
number of dimensions. More importantly, this approach yields the maximum
freedom to reorder operations in the innermost loop dimensions. 

%% file: table-results-summary.tex
\begin{table*}[!htb]
\caption{\label{tab:summary} Geometric Mean Circuit Depth and Added Gates across topologies, loop transformations and qubit allocators}
\center{
{\footnotesize
\begin{tabular}{|c|c|c|c|c|c|c|c|c|}
\hline
{\bf Topology x Transformation} & \multicolumn{4}{c|}{\bf Circuit Depth (No.Gates)} & \multicolumn{4}{c|}{\bf No. of Added Gates} \\
                                & {\bf jku} & {\bf ibm}  & {\bf sabre} & {\bf wpm} & {\bf jku} & {\bf ibm}  & {\bf sabre} & {\bf wpm} \\
\hline                                                               
\topomring ~x base   & 	204.79 & 	314.63 & 	208.17 & 	358.02             & 	317.25        & 	568.47 & 	290.30       & 	723.47\\
\topomring ~x feautrier   & 	211.31 & 	311.54 & 	203.04 & 	358.47       & 	327.48        & 	556.27 & 	275.04       & 	747.09\\
\topomring ~x plutomin   & 	203.85 & 	312.80 & 	204.96 & 	367.19         & 	316.87        & 	562.75 & 	269.52       & 	769.92\\
\topomring ~x plutomax   & 	206.43 & 	309.15 & 	205.16 & 	356.87         & 	321.21        & 	550.15 & 	284.55       & 	735.11\\
\hline                                                                   
\topogrid ~x base   & 	162.31 & 	228.26 & 	169.33 & 	307.15             &  225.17        & 	419.21 & 	{\bf 203.31} & 	584.08\\
\topogrid ~x feautrier   & 	167.22 & 	225.93 & 	169.20 & 	302.71         & 	{\bf 249.83}  & 	417.37 & 	{\bf 226.35} & 	626.60\\
\topogrid ~x plutomin   & 	161.93 & 	228.69 & 	165.52 & 	308.05         & 	225.12        & 	425.11 & 	208.55       & 	626.09\\
\topogrid ~x plutomax   & 	161.61 & 	228.71 & 	170.60 & 	298.12         & 	{\bf 224.40}  & 	415.32 & 	223.36       & 	590.60\\
\hline                                                                   
\topotiled ~x base   & 	191.55 & 	250.50 & 	198.37 & 	319.59             & 	336.20        & 	469.33 & 	288.63        & 	370.72\\
\topotiled ~x feautrier   & 	196.23 & 	254.10 & 	192.31 & 	321.07       & 	344.30        & 	473.35 & 	{\bf 294.37}  & 	385.64\\
\topotiled ~x plutomin   & 	190.22 & 	249.75 & 	188.44 & 	315.98         & 	330.93        & 	473.35 & 	{\bf 266.59}  & 	382.62\\
\topotiled ~x plutomax   & 	195.73 & 	246.92 & 	194.65 & 	329.98         & 	333.47        & 	461.80 & 	287.61        & 	379.69\\
\hline
\end{tabular}
}
}
\end{table*}

%% file: results.tex

\input{exp-general}


\label{sec:expsetup}
We have implemented the \axl~ language, and the analyses and transformations
described in Section \ref{sec:polyhedral} as a source-to-source compiler
toolchain using the Integer Set Library \cite{isl.2010}.  Quantum circuits are
written and compiled with \axl~ to produce the loop structure, which 
is then post-processed to construct a compilable C-program. This is then
compiled into an executable binary to finally generate the stream of quantum assembly
operations, targeting either the ProjectQ \cite{projectq.2018} compiler or
OpenQASM 2.0 \cite{openqasm.arxiv.2017}.  In particular, the results shown here
are obtained with QASM files.  The \axl~ benchmarks were compiled
on an 8-core AMD Ryzen 7 2700X - 2.1GHz with 105GB DRAM, 32KB L1, 512KB L2 and
8MB L3 cache.

The overall goal of our experimental evaluation is to demonstrate
that even ``classical'' high-level loop transformations,
i.e. not yet tailored to the quantum computing
domain, can synergistically work with back-end compiler
optimization such as Qubit Allocation. Thus, our goal is to determine
where (in which experimental configurations), why (interactions of loop transformations
with the qubit allocator)
and when (in some specific stage of a circuit) can impact tha qubit allocation
result, either for good (improved quality, i.e. shallower circuits) or 
bad (bigger circuits).

{\bf Testbed and Protocol:}
We use the Enfield compiler \cite{enfield-web,enfield-git}, which implements
several qubit allocation techniques. In particular, 
we consider four of the qubit allocation
algorithms covered in Sec.\ref{sec:background}: {\em sabre} \cite{sabre.asplos.2019},
{\em jku} \cite{jku.date.2018}, {\em wpm} \cite{siraichi.cgo.2018},
and {\em ibm} \cite{openqiskit.2018}. Enfield allows to collect
statistics such as allocation time, circuit depth (no. of gates in critical path), circuit size (total no. of gates)
among many others. The methodology we follow is the same as that of {\bf sabre} \cite{sabre.asplos.2019}
when comparing against {\bf jku}, while {\bf wpm} and {\bf jku} compare against the {\bf ibm} mapper
in their respective publications. As context, the {\bf jku} paper reported
a 23\% improvement (no. of elementary gates added) w.r.t. to {\bf ibm}, while {\bf sabre} reports an 
average improvement of 14\% over {\bf jku}, considering their qft and large benchmark
categories.

We use the circuit depth and size as primary quality metrics,
which vary with the allocator
due to the number of {\em quantum SWAP} and {\em REVERSE} operations
introduced to advance the program state, usually between adjacent layers.
Each of these operations is implemented with elementary gates.
Each SWAP is implemented with 3 CNOT, while each reverse operation
requires 4 Hadamard (H) and 1 CNOT gates.

Our experimental testbed consists of the eight quantum circuits listen
in Table \ref{tab:benchmarks}, which were taken from the
literature \footnote{Please refer to the Appendix for a larger \axl~ example
and circuit diagrams}.  
We generate four different variants for each benchmark:
pass-through mode performing only code generation (\trbase); the transformed
code obtained by applying the Feautrier scheduling algorithm (\trfeautrier)
\cite{feautrier.ijpp.1992a,feautrier.ijpp.1992b}; and the two well known
Pluto's fusion heuristics, maximal loop fusion (\trplutomax) and maximal loop
distribution (\trplutomin). We note that the \trbase~ variant, while only involving code generation,
can already be considered an optimized program, as the polyhedral scanning
process will produce minimal control overhead that can vastly differ
from hand-written loop-based code \cite{bastoul.pact.2004}. The latter two variants represent ends of the
classical locality spectrum.  Although Enfield already repeats the allocation
process for the algorithms with random properties (5 for {\bf sabre} and 20 for
{\bf ibm}), we still found substantial variation in the results. Thus, for each
allocator we repeat the allocation process 10 times and report their arithmetic
mean. We also include error bars showing the standard deviation.

\begin{figure}[h]
\centering
\subfloat[\topogrid]{\includegraphics[width=0.28\linewidth]{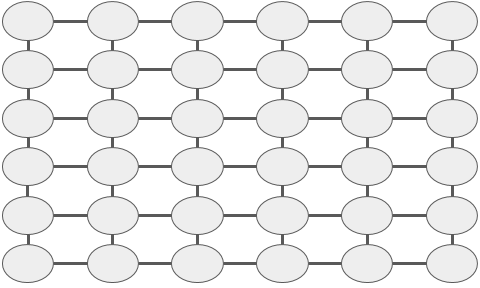}}
\hspace{1em}
\subfloat[\topomring]{\includegraphics[width=0.28\linewidth]{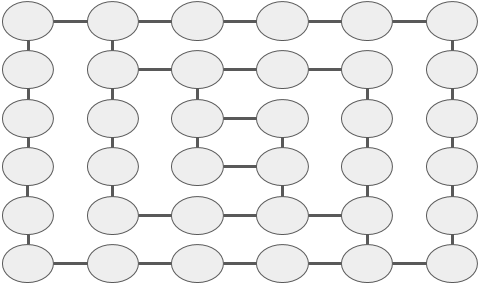}}
\hspace{1em}
\subfloat[\topotiled]{\includegraphics[width=0.28\linewidth]{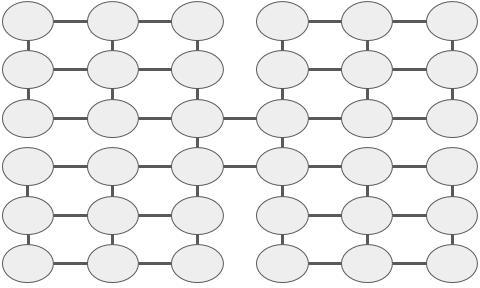}}
\caption{\label{fig:topologies}Evaluated coupling graphs (topologies)}
\end{figure}

{\bf Evaluated Topologies:}
Just as in classical computing the underlying architecture (e.g. multi-core CPU, many-core CPU, GPUs, etc)
can have a substantial impact on the program's performance, in quantum computing the quality
of the resulting qubit allocation can depend of the underlying architecture (topology).
We thus evaluate three coupling graphs  with the same number of qubits (36), and slightly vary
the graph's connectivity to produce different properties.
The three coupling graphs used are shown in Fig.\ref{fig:topologies}, and include
a 6$\times$6 grid (\topogrid), a graph with three doubly-linked concentric rings (\topomring),
and a four 3x3 tiled array of qubits (\topotiled).
All graphs have the same diameter (10), but differ in their bisection bandwidth, which is 6 for
\topomring~ and \topogrid, and only 2 for \topotiled. 
In addition, having more qubits with higher degree
enhances the architectural parallelism. For instance, if a given qubit has degree $d$, then
completing an update (write) on this qubit can enable up to $d-1$ gate-to-gate dependences.
For \topomring, \topotiled~ and \topogrid~ the number of gates with degree 3 or greater
are 8, 24 and 32, respectively. This design decision differs from 
the IBM QX20 (Tokyo) coupling graph, which varies over time, and which typically has
several qubits with degree 5 or 6. 

\begin{figure*}[!htb]
\includegraphics[width=0.49\linewidth]{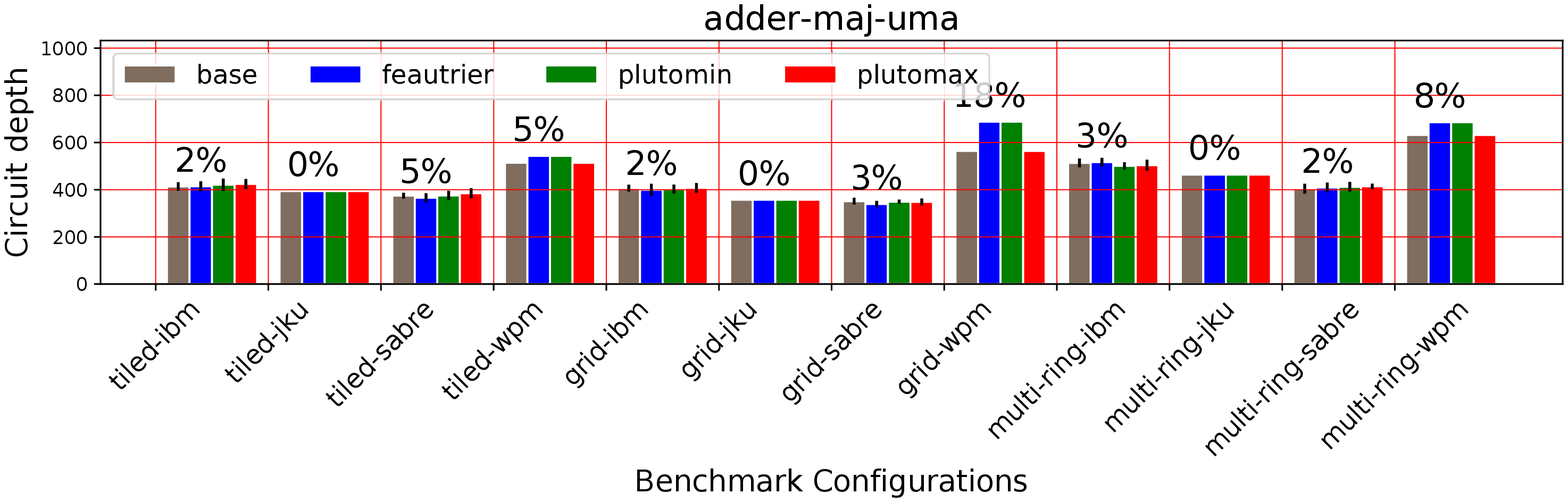}
\includegraphics[width=0.49\linewidth]{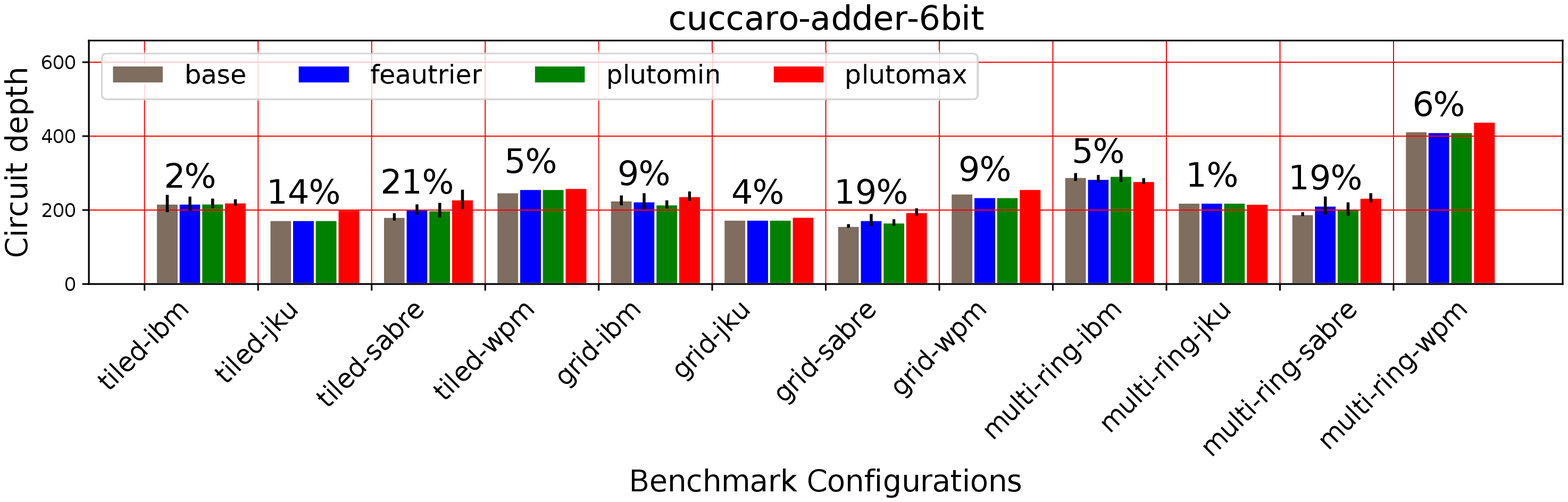}
\\
\includegraphics[width=0.49\linewidth]{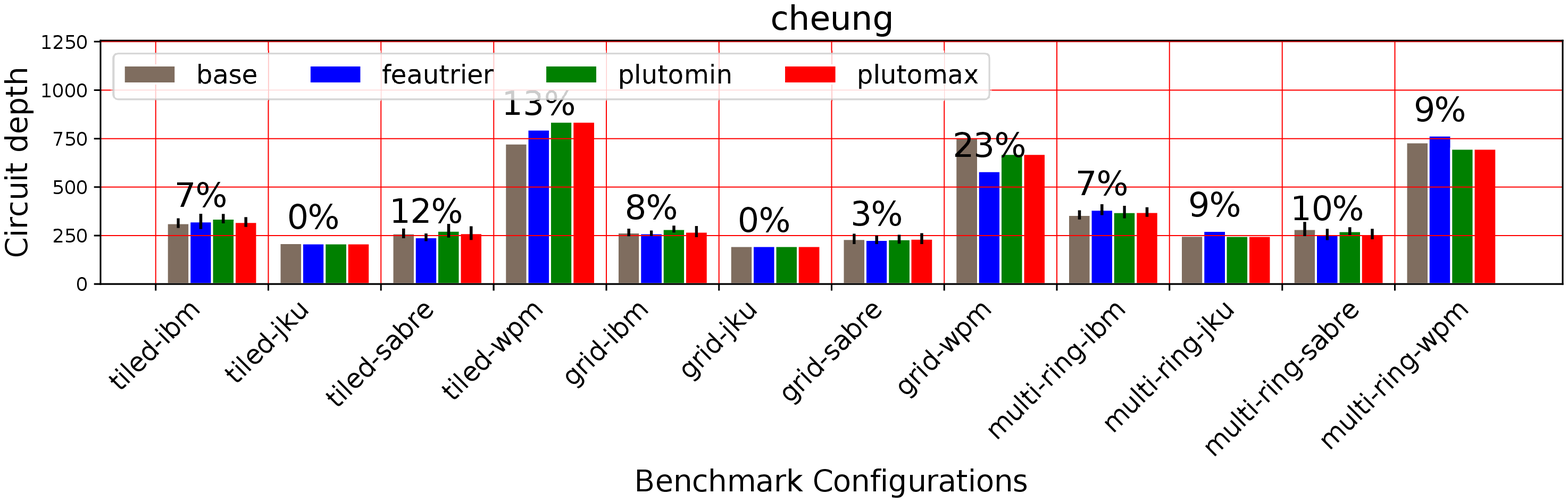}
\includegraphics[width=0.49\linewidth]{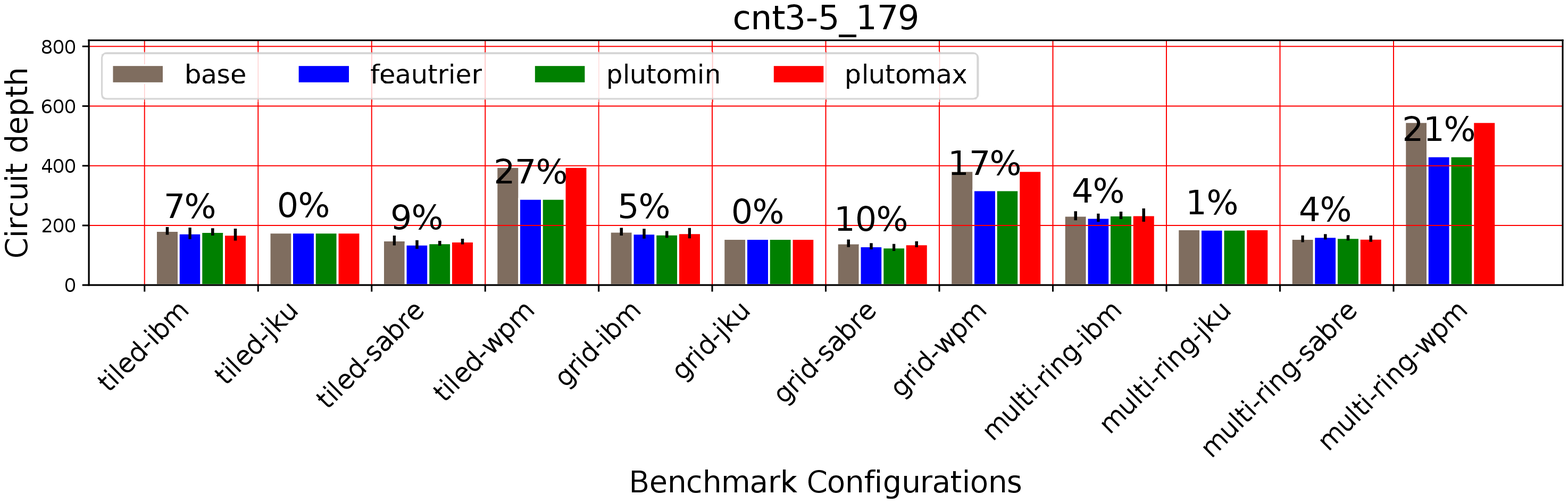}
\\
\includegraphics[width=0.49\linewidth]{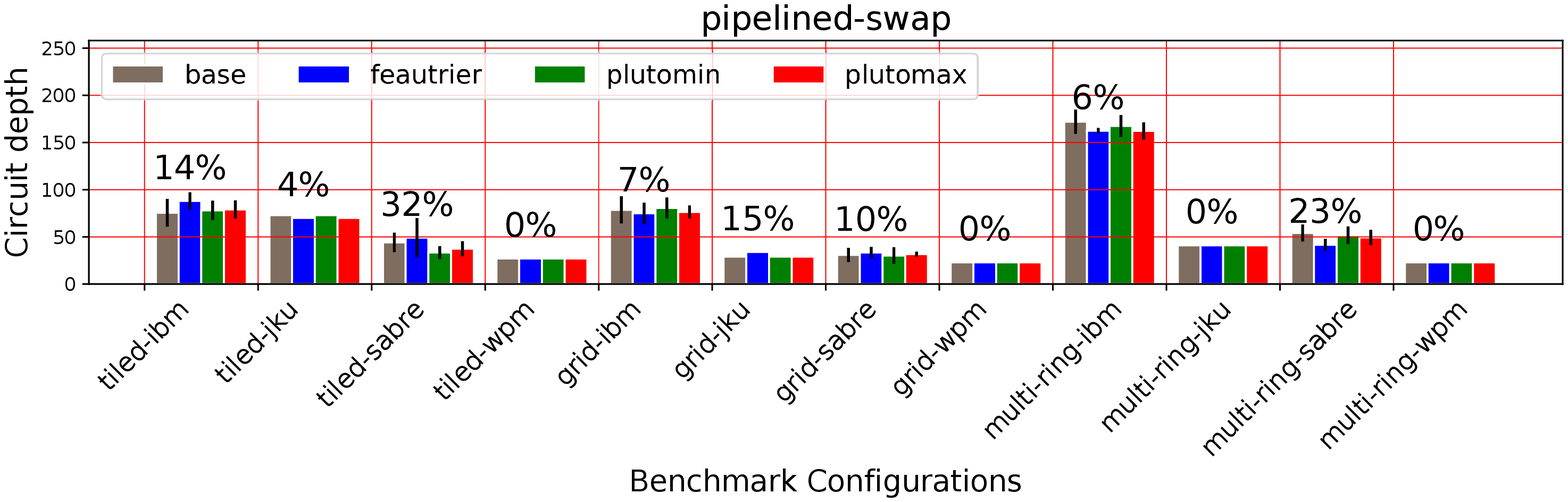}
\includegraphics[width=0.49\linewidth]{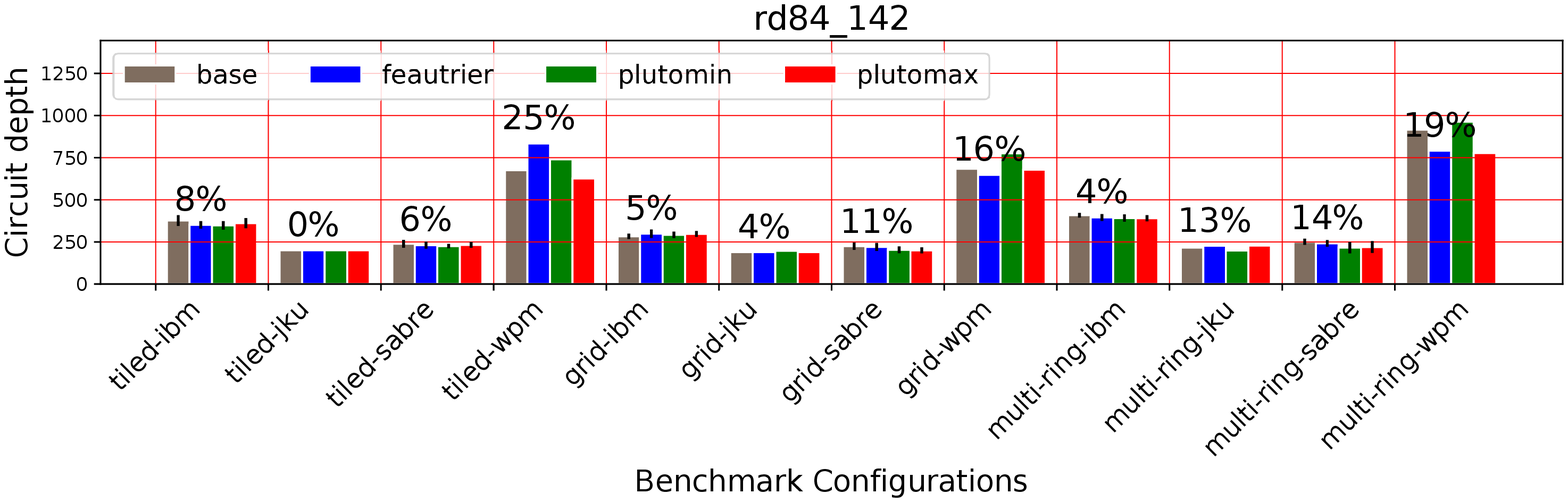}
\\
\includegraphics[width=0.49\linewidth]{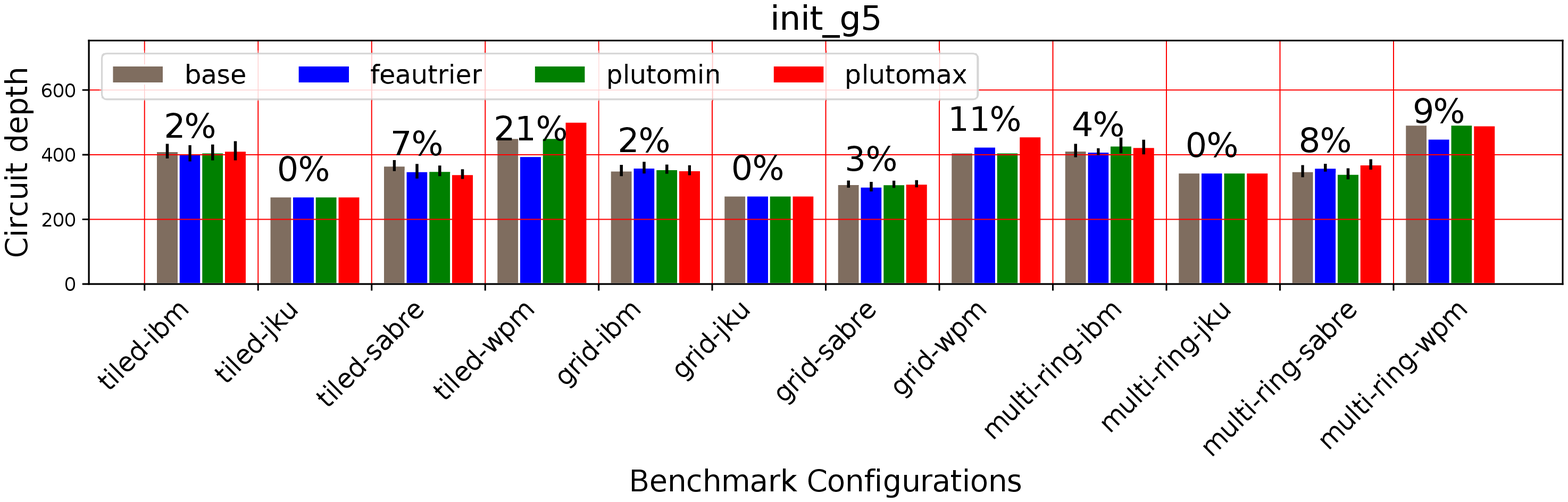}
\includegraphics[width=0.49\linewidth]{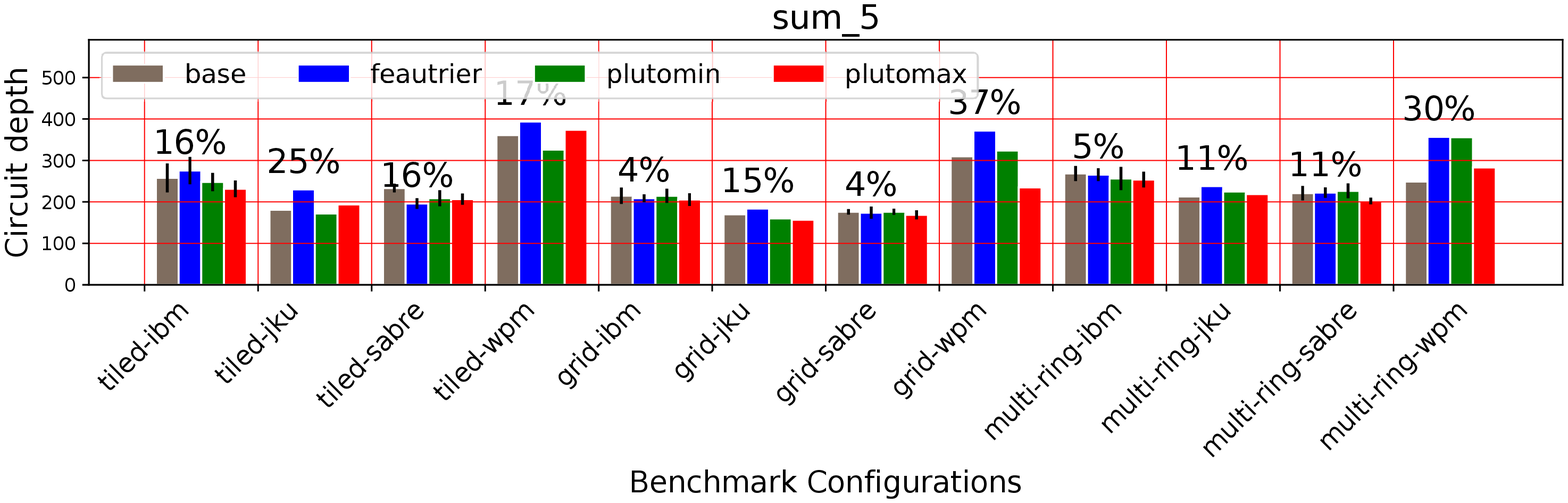}
\caption{\label{fig:all-topo}Impact of Affine Loop Transformations on circuit depth for \topogrid, \topomring~ and \topotiled}
\end{figure*}


{\bf TL;DR:}
Overall, we observe that quantum qubit allocation algorithms
and quantum programs can benefit from the power of
global cost functions offered by polyhedral
optimization techniques. Our evaluation shows that
even the most advanced allocators ({\bf jku} and {\bf sabre})
can improve up to 34\%, while achieving 60\% on {\bf wpm}.
We also observe and confirm general trends such as allocation variability
due to the inherent data- and pipelined- parallelism,
distance among qubit operands,
or the hardware parallelism available.

\input{results-summary}

\input{results-individual}

\input{results-pbs-scaling}

%% file: exp-general.tex

\begin{table*}[h]
\caption{\label{tab:benchmarks} Quantum Circuits Evaluated with Default Parameters}
\center{
\begin{tabular}{|c|c|c|c|c|c|c|c|}
\hline
{\bf Benchmark} & {\bf Source} & {\bf No.State-}  & {\bf No.Polyhedral} & {\bf Para-}  &  {\bf No.Qubits} & {\bf No.QASM}  & {\bf Output} \\
{\bf Name}      &              & {\bf ments}      & {\bf Dependences}    & {\bf meters} &                  & {\bf ops}      & {\bf Lines}  \\
\hline
\baddermau  & \cite{cuccaro.arxiv.2004} & 3   & 11   &   N=9        &   20    &   55      & 13-15 \\
\bcuccaro   & \cite{cuccaro.arxiv.2004} & 6   & 50   &   N=6        &   14    &   46      & 26-36 \\
\bsum       & \cite{Takahashi:qic.2008} & 7   & 41   &   N=5        &   11    &   36      & 22-32 \\
\binit      & \cite{Takahashi:qic.2008} & 8   & 33   &   N=5        &   11    &   35      & 24-34 \\
\bcheung    & \cite{cheung.wqccc.2008}  & 1   & 1    &   N=6        &   18    &   21      & 5     \\
\bpipelined & \cite{pipelined-swap}     & 13  & 28   &   N=6        &   14    &   75     & 37-43 \\
\bcnt       & \cite{revlib.2008}        & 2   & 14   &   N=5        &   19    &   30      & 14-20 \\
\brd        & \cite{revlib.2008}        & 3   & 36   &   M=2, N=4   &   15    &   28      & 16-32 \\
\hline
\end{tabular}
}
\end{table*}

%% file: results-summary.tex
\paragraph{General Trends}
We summarize in Tab.\ref{tab:summary} the geometric mean of the circuit depth (no.gates) and number 
of added gates across all combinations of topologies, affine transformations and qubit allocators.
The first trend to observe is that the depth of the circuit tends to increase with 
with lower architectural parallelism (e.g. no. of qubits with degree 3 or greater), as expected.
Next, we observe than even the state-of-the-art allocators, {\bf jku} and {\bf sabre}, can greatly
benefit from affine transformations. In particular, we observe up to a 10\% gap of added gates ($\max \Delta_{transform} / \max(added)$)
due to loop transformations in the \{\topogrid,{\bf jku}\}, \{\topogrid,{\bf sabre}\} and
\{\topotiled,{\bf sabre}\} configurations.
In regard to {\bf jku},
this qubit allocator produces the shortest circuits (depth-wise) in all three topologies when used in combination with \trplutomin,
and producing its worst circuit depth when combining it with \trfeautrier. This trend is nearly
the same for {\bf sabre}. Lastly, we don't observe any obvious trend
involving the {\bf ibm} allocator.

\begin{figure*}[!htb]
\includegraphics[width=0.70\linewidth]{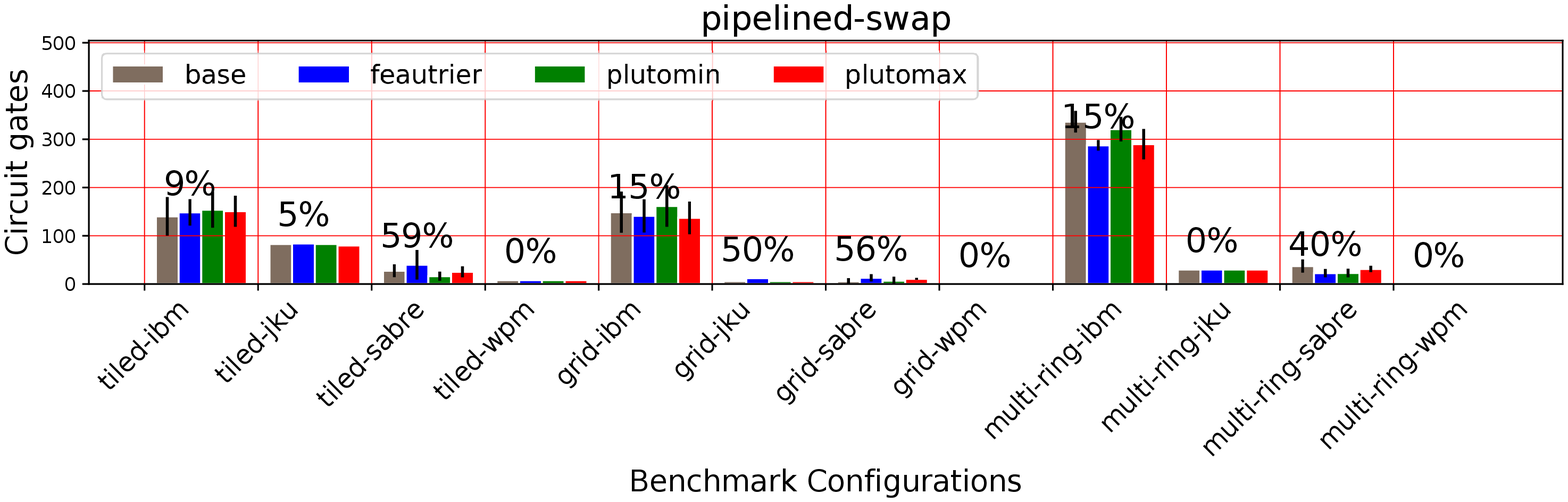}
\\
\includegraphics[width=0.70\linewidth]{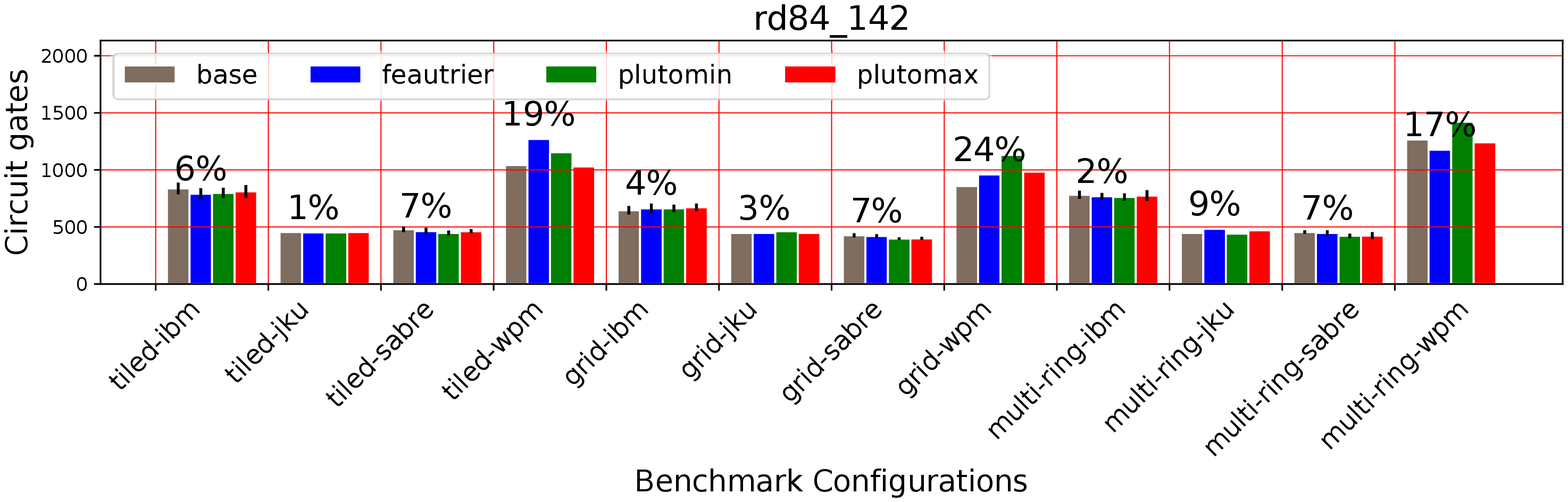}
\\
\includegraphics[width=0.70\linewidth]{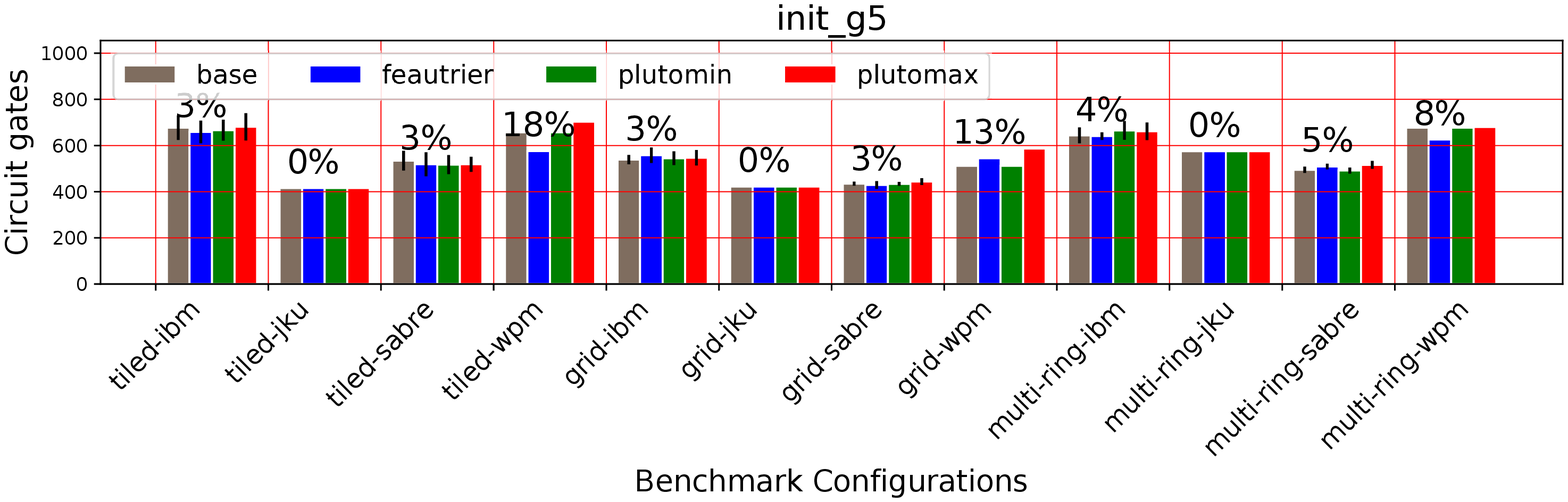}
\\
\includegraphics[width=0.70\linewidth]{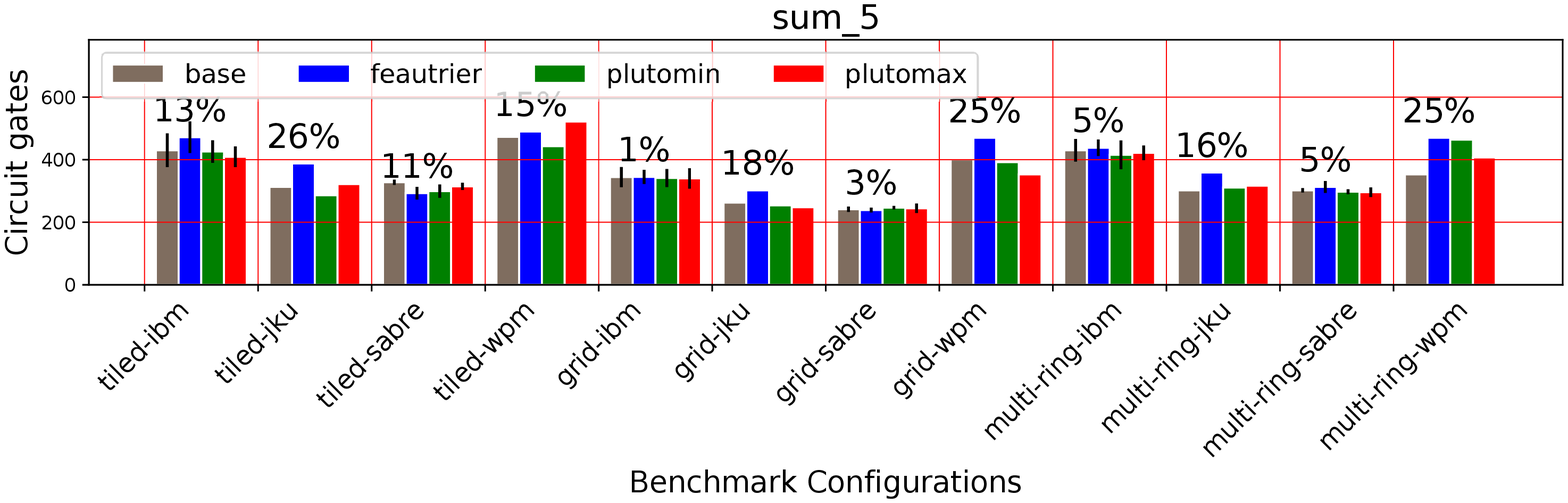}
\caption{\label{fig:all-topo-circ-size}Impact of Affine Loop Transformations on circuit size for \topogrid, \topomring~ and \topotiled}
\end{figure*}

In regard to the number of gates added by each allocator (last 4 columns), we observe that this metric
also correlates with the architectural parallelism of the topologies used.
We also observe that an overall increase of gate count does not necessarily correlate
with better circuit depth. This phenomenon can be seen when comparing the \{\topogrid, {\bf jku}\} and
\{\topogrid, {\bf sabre}\} with  \{\topomring, {\bf jku}\} and \{\topomring, {\bf sabre}\}.

%% file: results-individual.tex
\begin{figure*}[!htb]
\includegraphics[width=1.0\linewidth]{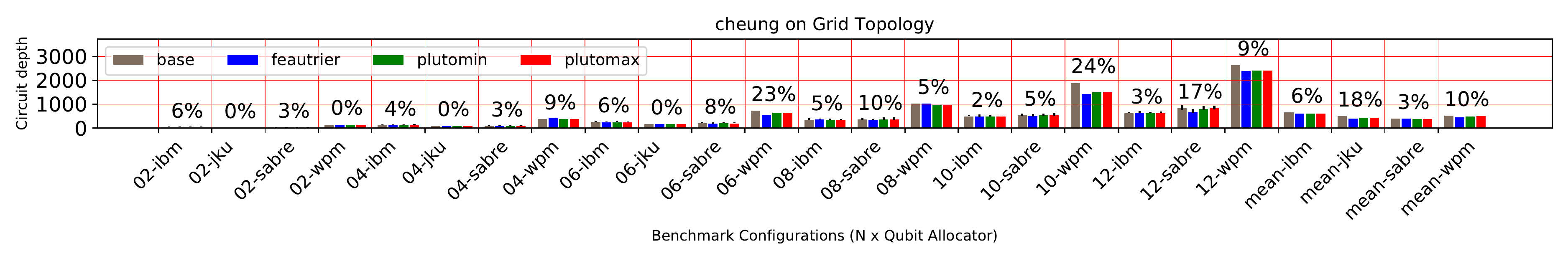}
\\
\includegraphics[width=1.0\linewidth]{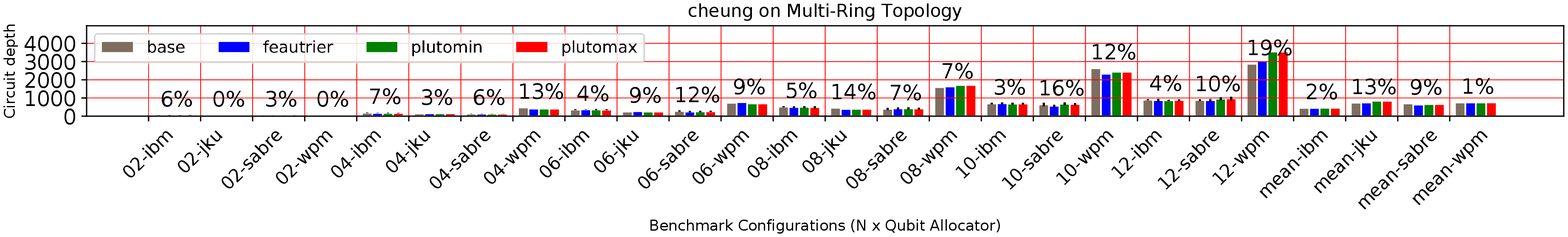}
\\
\includegraphics[width=1.0\linewidth]{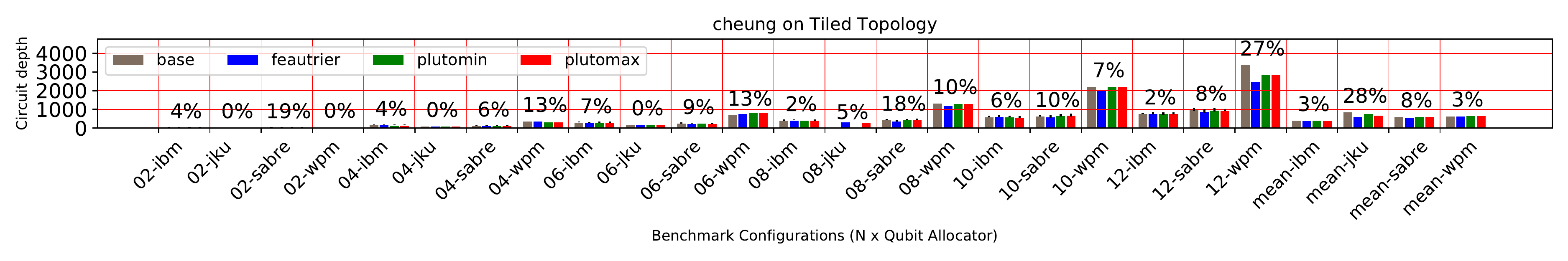}
\caption{\label{fig:scaling-pbs}Impact of problem size scaling on circuit depth for \bcheung~ circuit on topologies \topogrid~ (top), \topomring~ (middle) and \topotiled~ (bottom)}
\end{figure*}

\paragraph{Individual Analysis with Default Parameters}
We next dissect the behavior of our eight quantum circuits, and show the achieved circuit depth
in Fig.\ref{fig:all-topo}. For each circuit, we cluster the results by
{topology $\times$ allocator}. Each bar represents the mean of 10 repetitions, and include their
corresponding standard deviation. In addition, each cluster is also tagged with the highest
percentual variation between the highest and lowest depth ($(depth_{max} - depth_{min}) / depth_{max}$)
among the loop transformation for the same topology and allocator.
In general, we expect lower variation for the \topogrid~
topology than for the \topomring~ and \topotiled~ topologies.

Turning our attention to the \bpipelined~ circuit. This benchmark represents the swap between 
two distant qubits (e.g. $q_0$ and $q_{35}$). The circuit initially exhibits 2-way parallelism,
simultaneously starting with the two qubits, and morphing into pipelined parallelism
in its steady-state (See Appendix for circuit diagram). 
Each CNOT operation being the input dependence to the operations on two adjacent qubits.
The interesting result here is that {\bf wpm}, which typical produces lower quality mappings,
consistently produces shallower circuits. We also observe up to a 23\% gap for {\bf sabre} on 
the \topomring~ topology yielded by the \trfeautrier~ transformation.

The \bcheung~ circuit exhibits N-way data-parallelism, N being the number of pipelined CCNOT (Controlled-Controlled NOT) gates found
at the bottom of the circuit. The first CCNOT  of each qubit can be mapped to a distinct
qubit to maximize parallelism and reduce the circuit depth. Intuitively, both the \trfeautrier~ and \trplutomax~ should yield
the highest benefit in terms of circuit depth, since maximizing the number of satisfied dependences 
per schedule dimension equates (for this benchmark) to minimizing the maximal-dependence distance.
The circuit's depth drastically increases with the quality of the qubit
allocator. This is due to the fact that the control qubits for each operation start the closest and
separate as the state evolves. This, in turn, requires more swaps operations to make the qubit
operands adjacent. 

The arithmetic adder using the majority and unmajority circuit pattern, \baddermau, exhibits
a highly serial form in its steady-state. The only opportunities for reordering transformations
arise from the potential fusion/distribution of the CNOT operations with the sequence of CCNOTs
following later. In general terms, we expect the \trplutomax~ to have the highest impact on the circuit depth.
The same observations hold for the \bcuccaro~ adder, which differs from the former in the
degree of data-parallelism. This translates to much shallower circuits, roughly half the depth of
the \baddermau~ counterparts. It also differs in that \trbase~ and \trplutomin~ typically achieve
the best circuit depth due to the already available parallelism. All four qubit allocators also
benefit from the higher scheduling and placement flexibility of \bcuccaro~ due to the lower number
of CCNOTs (3-operand gates) as well as the presence of CNOT and NOT gates.

\begin{figure*}[!htb]
\includegraphics[width=1.0\linewidth]{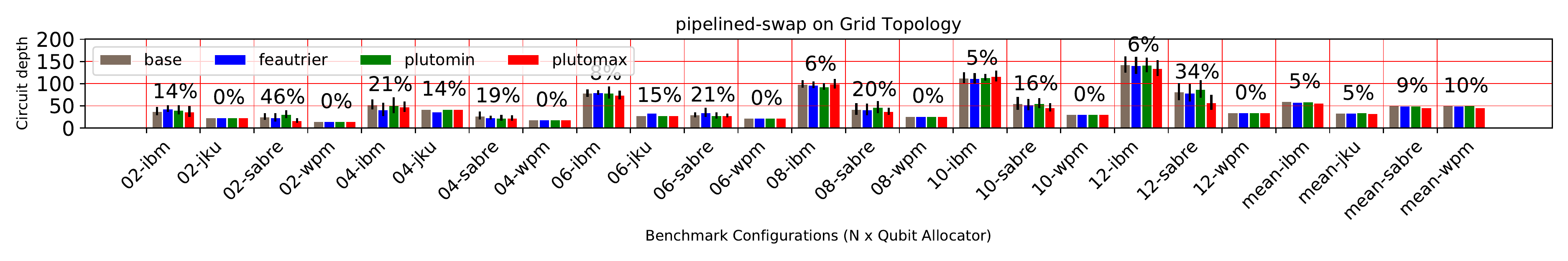}
\\
\includegraphics[width=1.0\linewidth]{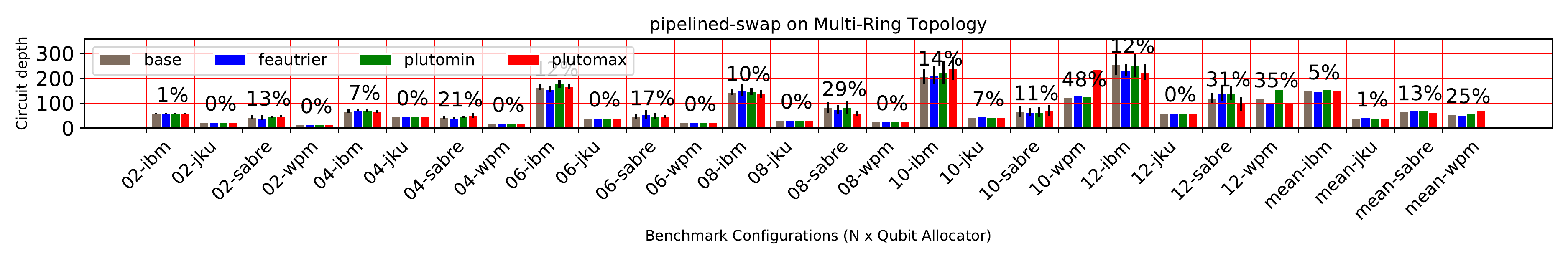}
\\
\includegraphics[width=1.0\linewidth]{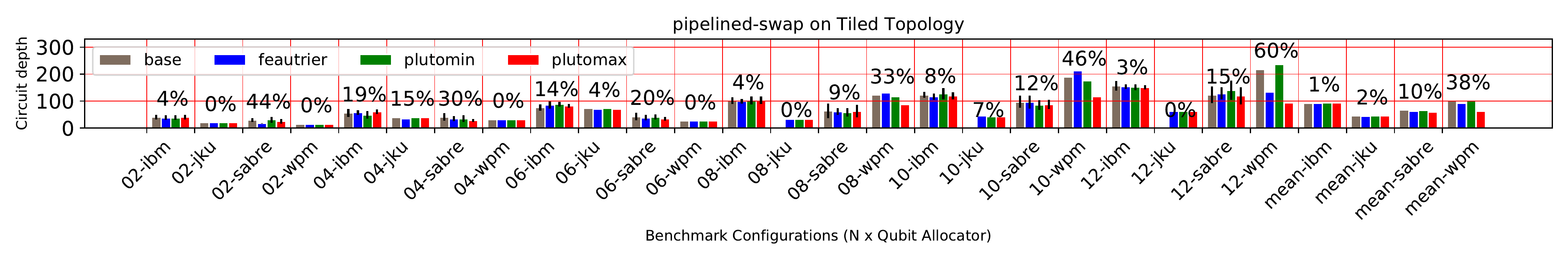}
\caption{\label{fig:scaling-pbs}Impact of problem size scaling on circuit depth for \bpipelined~ circuit on topologies \topogrid~ (top), \topomring~ (middle) and \topotiled~ (bottom)}
\end{figure*}

Benchmarks \bsum~ and \binit~ \cite{Takahashi:qic.2008} share similar properties. As both
exhibit distinct and noticeable parallel and serial phases. For instance, \bsum~'s initial 
and \binit~ ending phase
offer a lot of reordering freedom in the scheduling and placement of the CNOT operations.
In contrast, the middle and ending phases of \bsum~ and initial and middle ones of
\binit~ are mostly serial with stints of pipelined parallelism. In addition, the large and varying separation
between the gate operands induce a higher number of required SWAP operations. These combination
of factors make harder the prediction of the most suitable affine transformation, making the outcome
highly variable w.r.t to the coupling graph and the qubit allocator. So the overall topological
trends remain.

Circuit \bcnt~ also shares some similar features with circuit \bcuccaro. They both have a mix of data-
and pipelined parallelism, with short constant distance (3 or lower) among the gate operands.
These traits contrast with benchmark \brd~ which exhibits much larger, but constant separation (5).
\bcnt~ benefits from a slighter higher degree of data parallelism (left-most CNOTs).
Given the combination of characteristics, we expect \trplutomax~ to be more beneficial 
for \brd~, while \trfeautrier~ and \trplutomin~ to perform best (or close to) for \bcnt, 
as loop fission will effectively isolate the parallel operations from the more serial/pipelined ones. 
Empirically, both of these observations and hypotheses hold.

{\bf Impact on Circuit Size}
Next, we show in Fig.\ref{fig:all-topo-circ-size} the gap in circuit size for four of the previous
circuits. The analysis and justification is as before, revolving around the variability
in data- and pipelined parallelism, specific properties of the circuits such as the distance
between its qubit operands (fixed-short, fixed-large or variable), and the underlying topology.
In particular, we highlight that {\bf wpm} effectively detected that the {\bf pipelined-swap}
was a swap circuit in itself, unlike all other allocators, which still attempted to optimize
the circuit. We also observe circuit gaps ($\max \Delta_{transform} / \max(added)$) of up to 26\%
even for {\bf jku}, on \bsum.

%% file: results-pbs-scaling.tex
\paragraph{Impact of Circuit Scaling}

To complement our study, we perform a scaling evaluation of circuits 
\bpipelined~
and 
\bcheung. 
We focus on these benchmarks due to their high regularity and lack of singleton quantum statements
that require more specific scheduling. We vary the SCoP parameter N $\in \{2..12\}$. Beyond this,
the quantum register size is exceeded. To judge the overall scaling behavior, the last four
bar clusters in each figure show the arithmetic mean of each allocator varying the loop
transformation. We note that \bpipelined~ using {\bf jku} on the \topogrid~ topology only scaled up to N=6 (14 qubits),
at which point each run started taking above an hour to complete. Furthermore, at N=10, memory was being
exhausted in our benchmarking server. The time limit was also exceeded for the \trbase~ variant of \bpipelined~ using {\bf jku} 
on the other two topologies, for N $\geq$ 8, A consequence of the $A*$ search used by {\bf jku}. 
Next, we center our attention on the \bcheung~ benchmark,
which only utilizes CCNOT gates. This trait makes it fare best on topologies with higher number
of vertices with degree 3 or higher, i.e. \topotiled~(24) and \topogrid~ (32). The impact of
this requirement is notorious on the \topomring~(8) graph, which for N=4 requires a total
of 8 (4 producing and 4 consuming) gates. Past this point the number of swaps operations
introduced by the {\bf wpm} allocator nearly doubles for every increment of 2 on N. Given these resource
restrictions, we now refine our first assessment on this circuit, and expect the \trfeautrier~ transformation to fare best, as it was conceived
as a minimum latency, resource conscious transformation.
Equivalently, we also now expect \trplutomax~ to induce typically higher circuit depth.
The intuition here (in classical loop transformation terms) is that the \trplutomax~ heuristic would amount to an 
outer parallel loop and an inner serial loop, while the \trfeautrier~ transformation would 
make the outer loop serial, thereby exposing more inner parallelism. The general effect of this
trade-off for most qubit allocators is to expose earlier the gate-to-gate dependences. In particular,
this is most beneficial when the allocator makes local decisions using sliding windows.

Changing our focus to the \bpipelined~ circuit, we first note that, unlike the previous benchmark,
this one only consists of CNOT gates. Thus, the number of vertices with degree 3 or higher is
not a limiting factor. This provides more freedom to the qubit allocators. 
In general, the {\bf jku} allocator produces the circuits with the shallowest depth,
outperforming {\bf sabre} in every topology. Nonetheless, 
we highlight that {\bf wpm} in tandem with \trplutomax~ yields results comparable, and at times
better, than {\bf sabre}. This is relevant because {\bf wpm} is $10\times$ faster than {\bf sabre}
and $10\times$ to $15\times$ faster than {\bf jku}. The much improved circuit depth w.r.t
to {\bf wpm}'s impact on other benchmarks is due to an specific feature of \bpipelined~. 
If we divide the circuit into four quadrants, the north-east (NE), north-west (NW), 
south-east (SE) and south-west (SW), and consider them as the four statements to schedule,
we can notice that the NW and SW statements converge in the middle of the circuit. This means
that the underlying qubit allocator would benefit from finding those operations, the ones 
in the middle of the four quadrants, concentrated into a small window of operations. That is
precisely the effect of using \trplutomax~, and leads up to a 60\% circuit depth improvement 
when using {\bf wpm} on the \topotiled~ topology.

%% file: related.tex
\input{related-quantum-comp}

%% file: related-quantum-comp.tex
Modeling affine quantum circuits shares some similarities with polyhedral
process networks (PPN) \cite{sven.ppn.2009,nadezhkin.tecs.2013},
independent processes that communicate with unbounded FIFOs.
However, most of their work has focused on translating serial programs
into parallel hardware. Polyhedral and affine transformations have also successfully targeted
several architectures, among them
CPUs \cite{irigoin.popl.1988,lim.ics.1999}, GPUs \cite{baskaran.ics.2008,grosser.ics.2016,leung.gpgpu.2010,ppcg.taco.2013},
and FPGAs \cite{pouchet.fpga.2013,alias.fpga}. 
Functional languages have been proposed as viable candidates for the
specification of quantum programs: 
\cite{lapets.functionaldsls.2013} introduced a statically typed functional DSL;
\cite{altenkirch.lics.2005} introduced the QML language
focused on allowing the specification of reversible and irreversible quantum
computations and combining it with first order strict linear logic. 
The Scaffold language and the ScaffCC compiler 
\cite{scaffold.tr.2012,scaffcc.parco.2015}
allow for a modular organization of quantum
programs, and are equipped with control-flow constructs that allow  to
manipulate quantum gates.
Loke, Wang and Chen \cite{qcompiler.cpc.2013,loke.cpc.2016} developed the Qcompiler and
OptQC, which focused on the optimization of circuits by determining permutation
matrices that minimized the number of required swap gates. Their algorithm used
simulated annealing to determine near optimal number of swap gates. Svore et
al. developed Q\# \cite{qsharp.rwdslw.2018}, a DSL with a rich type system,
modular definitions, reversible operations, control-flow
constructs and qubit management. 
Similarly, quantum instructions sets \cite{smith.arxiv.2016} 
and assembly languages 
such as OpenQASM \cite{openqasm.arxiv.2017} and cQASM \cite{qasm.2017}
(the common Quantum Assembly language) have also been proposed.
QISKit \cite{openqiskit.2018} is an open sourced quantum toolkit,
available as a Python package, which
enables users to write programs with OpenQASM and run them in the 
IBM Quantum Experience \cite{ibm-quantum-experience.nature.2017}, 
a cloud service. These efforts embody
important software building blocks that new quantum compiler infrastructures
can build upon to develop more scalable and high-level frameworks.  RevKit
\cite{revkit.2012,revkit.2011} has also been used to support fully automatic
synthesis of quantum circuits \cite{soeken.date.2018}.
Quipper \cite{quipper.pldi.13,green.icrc.2013}, an embedded quantum 
programming language  for circuit specification developed by Green et al., 
introduced several features such as ancilla scope and reuse, classical to 
quantum circuit  lifting for automatic generation of application specific 
oracles, basic data types, boxed/procedural definition and reversing 
operators for defined circuits.
LIQUi|> 
\cite{liquid.arxiv.2014},
a DSL for quantum computing,
proposed language features such as static typing, 
opaque types for qubit and kets representation, and introspection
functionality that uses Microsoft's F\# language and .NET support.
\cite{qwire.popl.2017} introduced QWIRE, a language designed for
the specification of quantum circuits with strong type system and
safe properties for well defined circuits. In QWIRE, circuits are first-class
citizens, and provides boxing and unboxing functionality that enables
the composition of circuits. In addition, it also leverages dynamic
lifting to convert a quantum circuit to its classical equivalent.
Lastly, efforts to produce more robust quantum program
mappings for NISQ (Noisy Intermediate-Scale Quantum) architectures
by exploiting calibration parameters, scalability
and routing options are also being explored \cite{murali.asplos.2019}.


%% file: conclusion.tex
In this paper we have introduced the first polyhedral
quantum specification language and compiler, \axl.
We have demonstrated how off-the-shelf polyhedral
analyses and transformations from the classical HPC
world can be applied and beneficial to quantum computing.
We have found that even not-yet-tailored transformations
can improve state-of-the-art qubit allocators
such as {\bf jku} and {\bf sabre} by as much as 36\%,
and others such as {\bf wpm} by 60\%.
Clearly, much
work remains. An obvious follow up is to devise
different model-driven optimizations that 
embed parallelism constraints, as well as considering
the underlying machine topology.

%% file: appendix.tex

Due to space constraints and for better viewing, we include the set of
evaluated circuits in this appendix.  The set of circuit benchmarks evaluated
in our work are summarized in Tab.\ref{tab:benchmarks}.  These should be viewed
as small computational building blocks used in larger applications, in similar
spirit to modern and classic computational kernels such as \texttt{DGEMM}.

\begin{figure}[htb]
\centering
\subfloat[\binit~\cite{Takahashi:qic.2008}]{\includegraphics[width=0.75\linewidth]{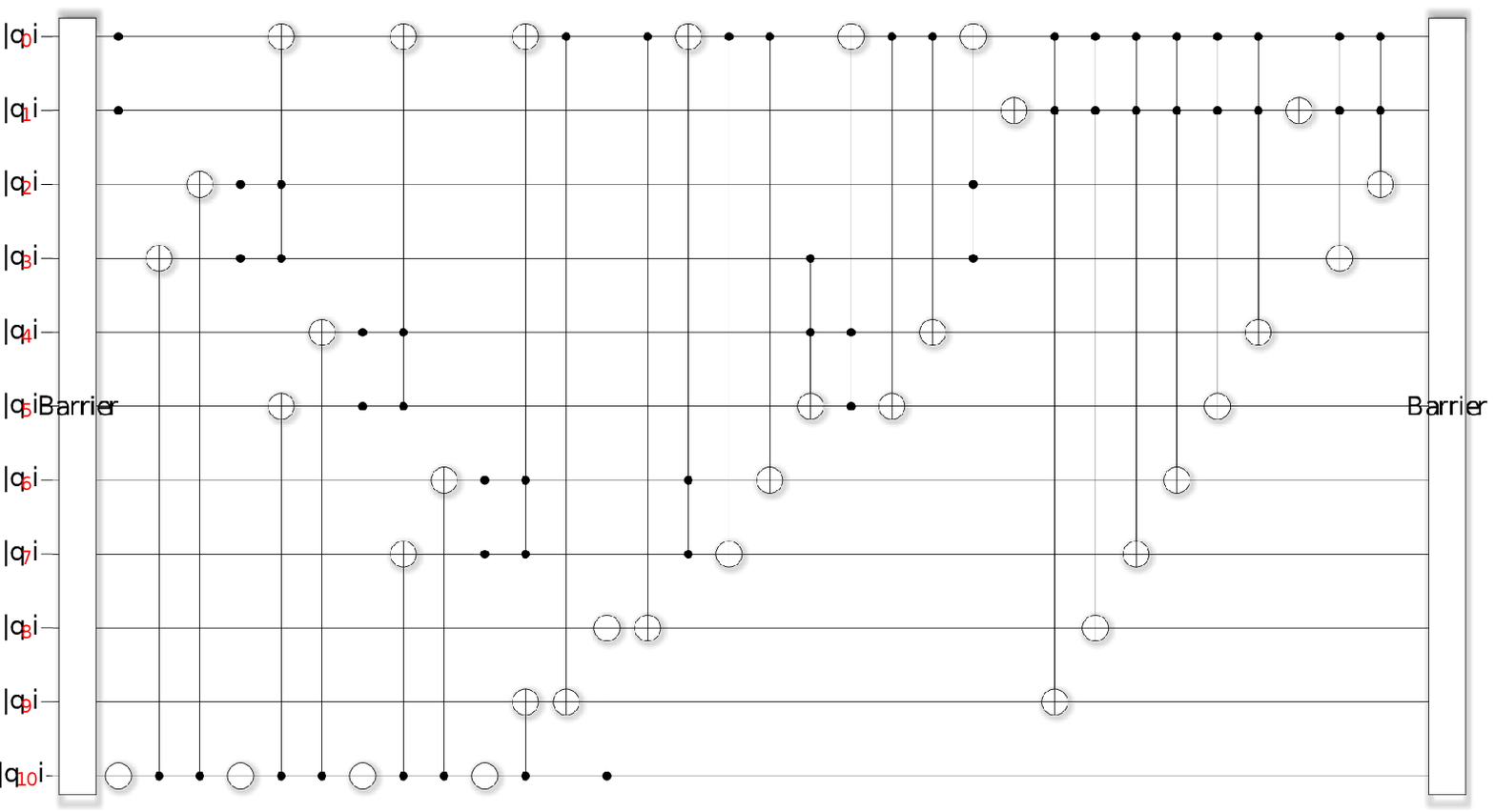}}\\
\subfloat[\bsum~\cite{Takahashi:qic.2008}]{\includegraphics[width=0.75\linewidth]{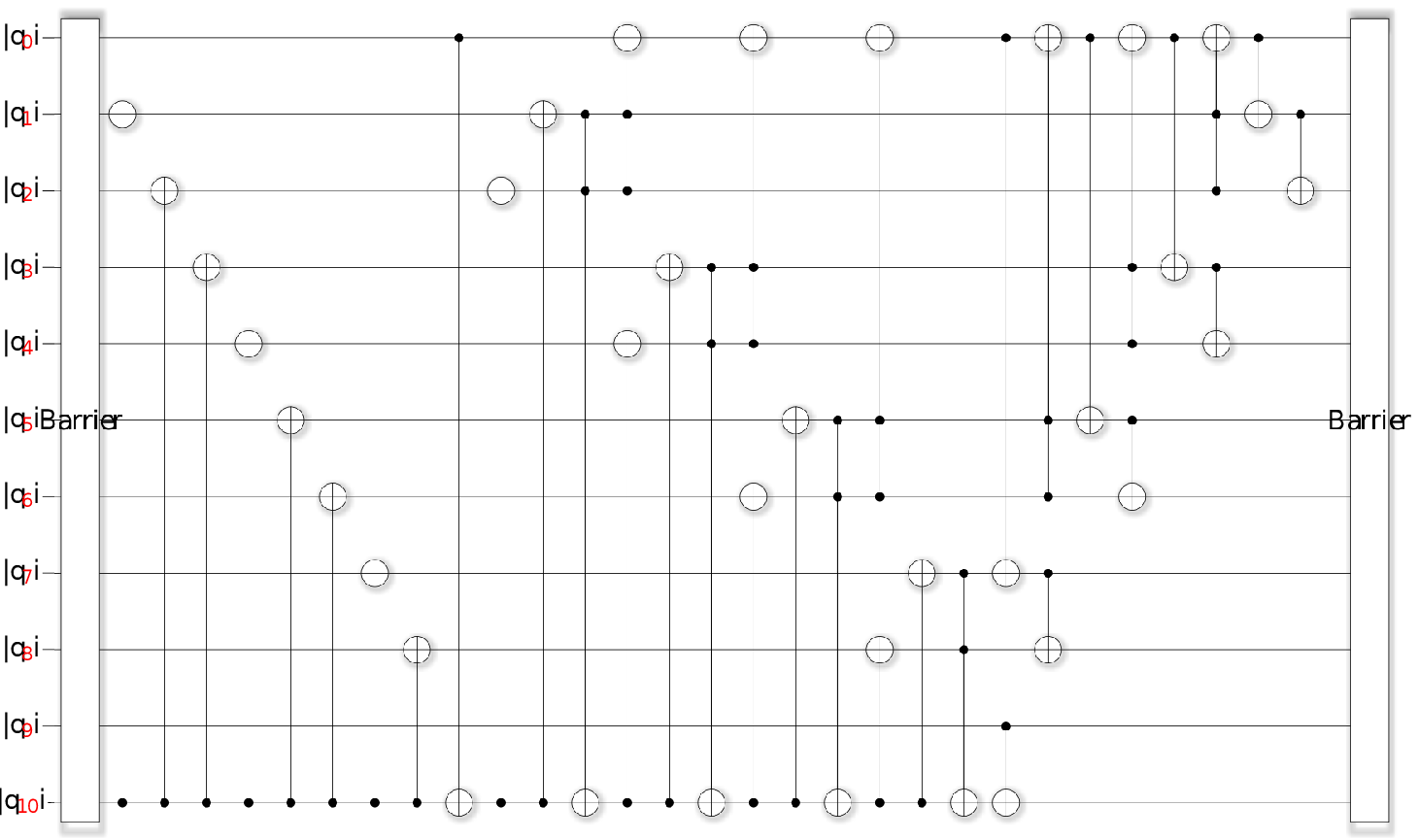}}
\caption{\label{fig:taka}\binit~ and \bsum~ circuits}
\end{figure}

\begin{figure}[htb]
\centering
{\includegraphics[width=0.58\linewidth]{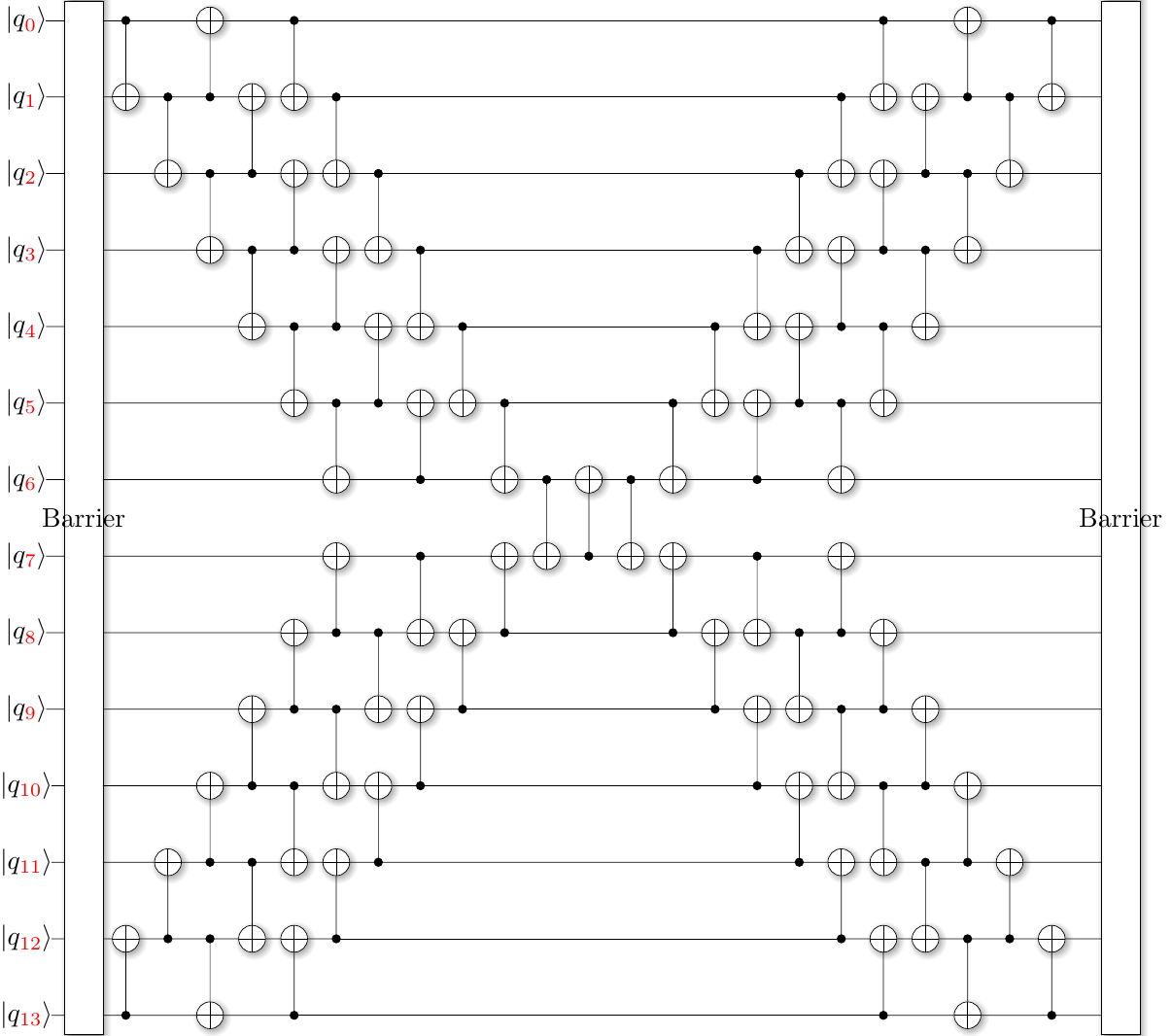}}
\caption{\label{fig:pipe}\bpipelined~\cite{pipelined-swap} }
\end{figure}

\begin{figure}[htb]
\centering
\subfloat[\bcnt~\cite{revlib.2008}]{\includegraphics[width=0.45\linewidth]{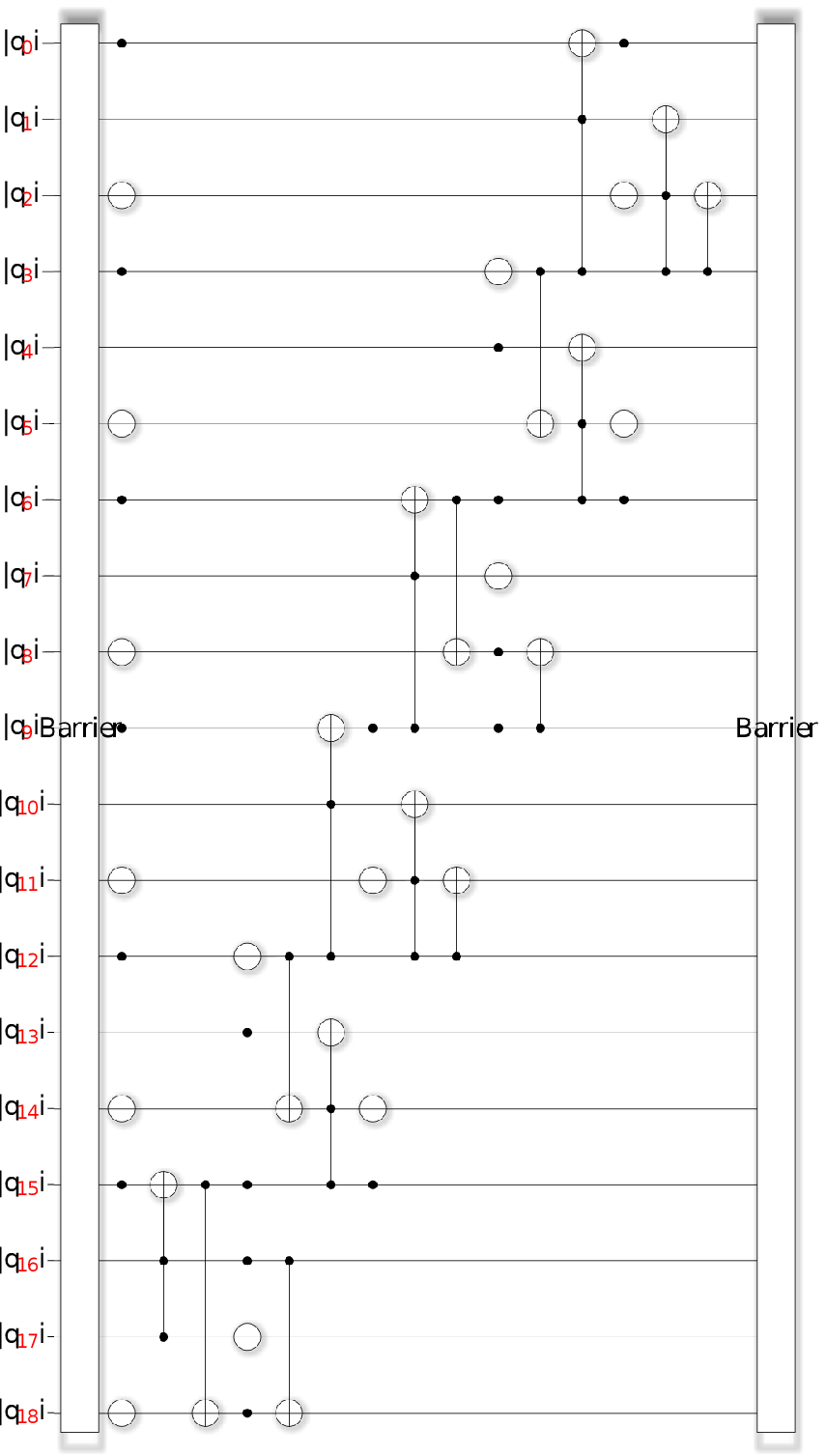}}
\subfloat[\bcheung~\cite{cheung.wqccc.2008} ]{\includegraphics[width=0.5\linewidth]{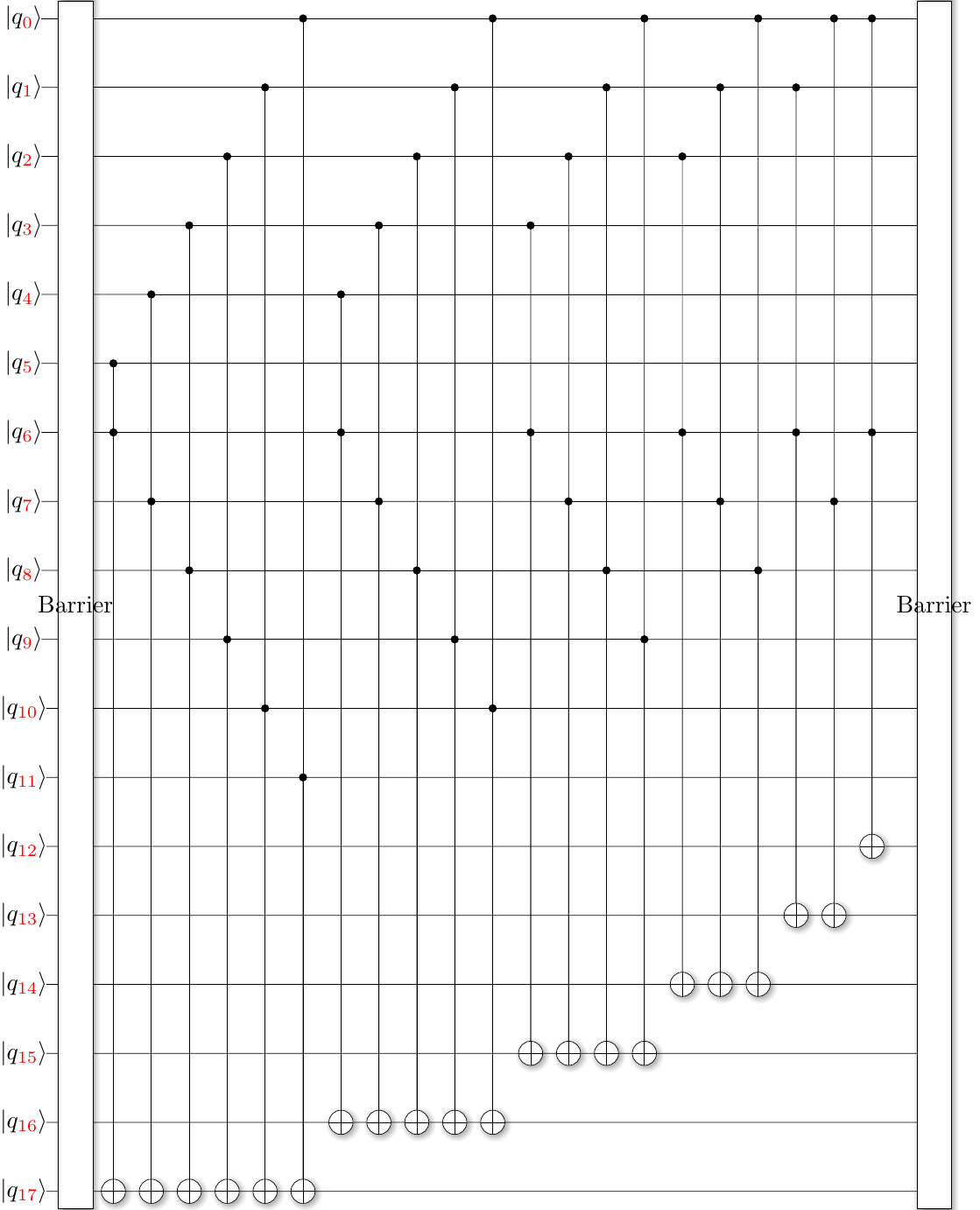}}
\caption{\label{fig:cheung}\bcnt~ and \bcheung~ circuits}
\end{figure}

\begin{figure}[htb]
\centering
{\includegraphics[width=0.65\linewidth]{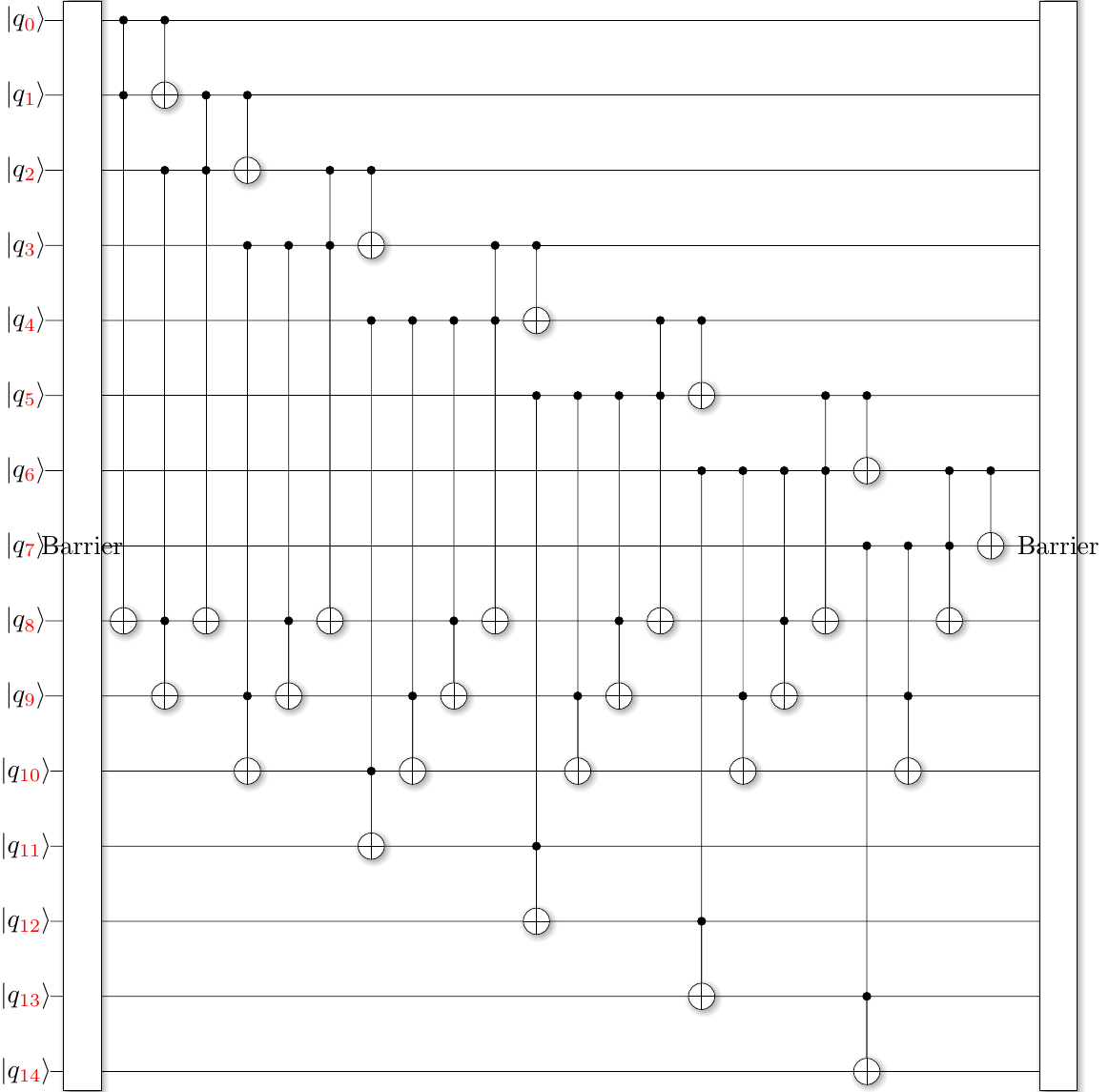}}
\caption{\label{fig:rd84}\brd~\cite{revlib.2008}}
\end{figure}

\begin{figure}[htb]
\centering
\subfloat[\baddermau~\cite{cuccaro.arxiv.2004}, N=6]{\includegraphics[width=0.8\linewidth]{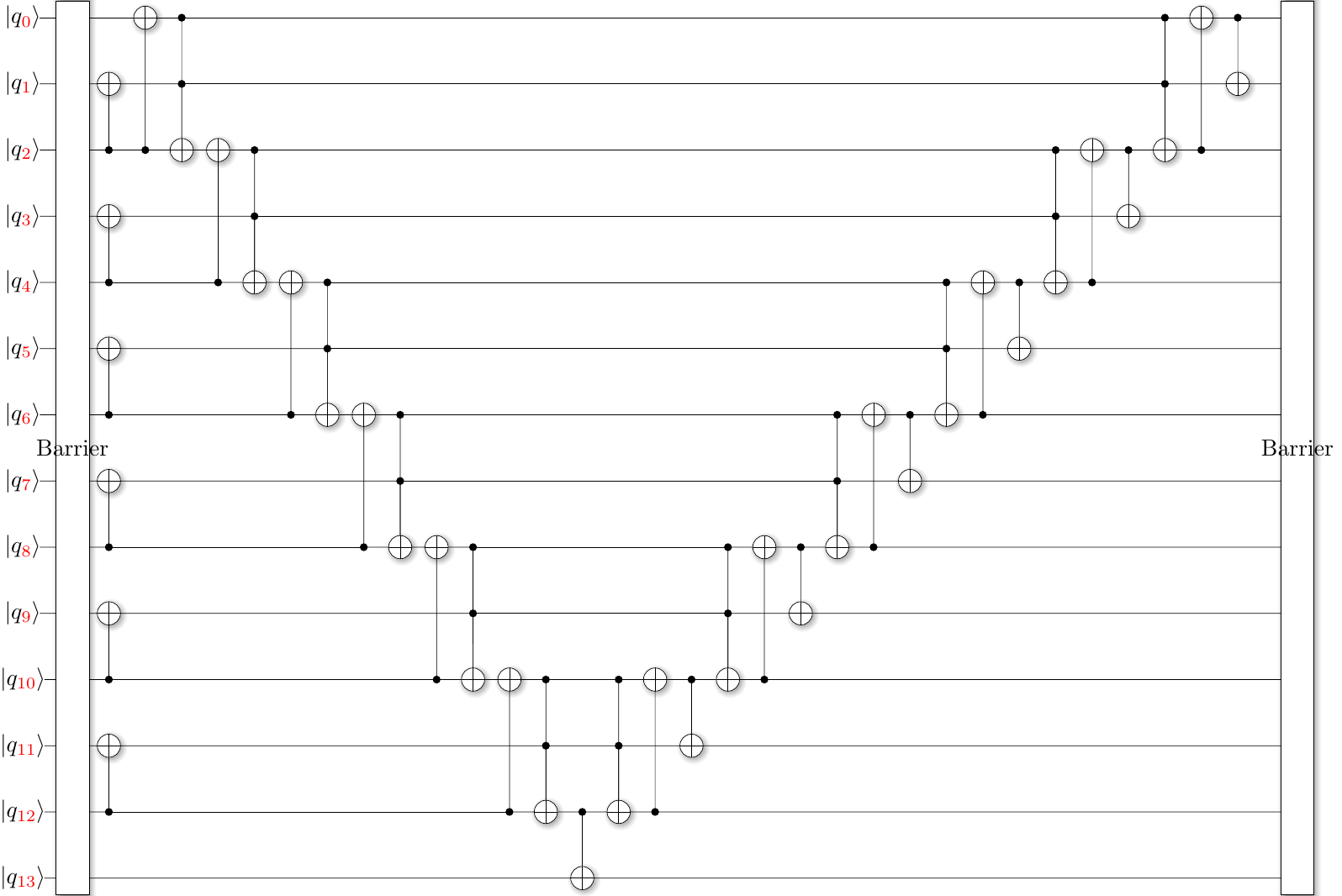}}\\
\subfloat[\bcuccaro~\cite{cuccaro.arxiv.2004}]{\includegraphics[width=0.6\linewidth]{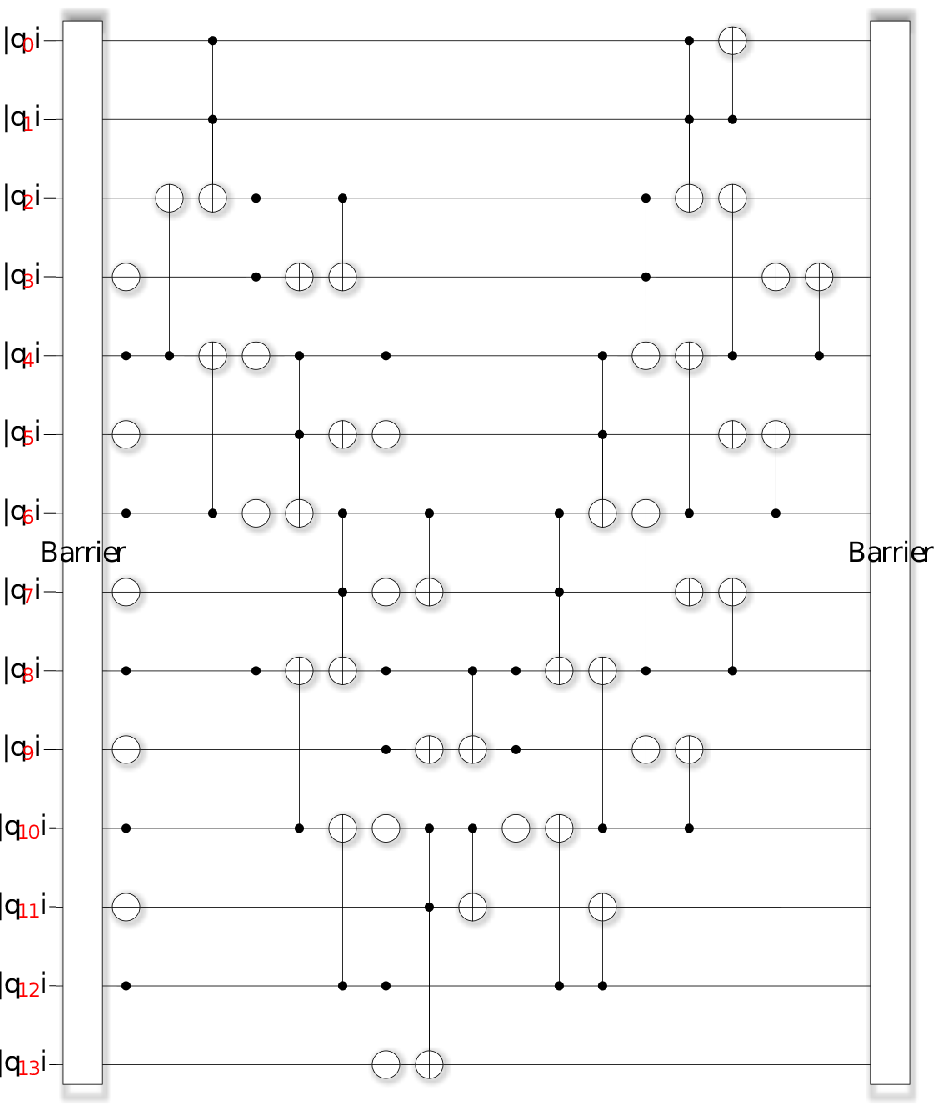}}
\caption{\label{fig:cuccaro}\baddermau~ and \bcuccaro~ circuits}
\end{figure}

Here we quickly summarize their overall role/goal:

\begin{itemize}

\item
{\bsum,\binit:} Sub-circuits used in \cite{Takahashi:qic.2008} corresponding to the computation
of initial
values and sum of an adder. The circuits do not use ancillary qubits, and have depth $O(n)$.
The corresponding circuit diagrams shown in Fig.\ref{fig:taka}.

\item
\input{pipelined-axl-source}
{\bpipelined:} Performs a qubit SWAP operation between two qubits $2N$ lanes apart.
We show in Fig.\ref{lst:axl-pipelined} its implementation in \axl, and its circuit
diagram in Fig.\ref{fig:pipe}.

\item
{\bcnt:} A 5 digit binary coded ternary counter with count control input (cnt) \cite{revlib.2008}.
Circuit diagram shown in Fig.\ref{fig:cheung}.

\item
{\bcheung:} sub-circuit used in computing the $\vec{d}$ component of 
the Elliptic Curve Discrete Logarithm problem (ECDLP) \cite{cheung.wqccc.2008}.
Circuit diagram shown in Fig.\ref{fig:cheung}.

\item
{\brd:} Counts the number of ones in the input\cite{revlib.2008}.
Circuit diagram shown in Fig.\ref{fig:rd84}.

\item
{\baddermau:} is a ripple-carry adder using the in-place ``MAJority'' (MAJ) and ``UnMajority and Add'' (UMA)
patterns \cite{cuccaro.arxiv.2004}. Circuit diagram shown in Fig.\ref{fig:cuccaro}.

\item
{\bcuccaro:}
Depth optimized ripple-carry adder of depth $2n+4$, with $2n-1$ time slices and $5$
CNOT time-slices \cite{cuccaro.arxiv.2004}. Circuit diagram shown in Fig.\ref{fig:cuccaro}.

\end{itemize}

%% file: pipelined-axl-source.tex
\begin{figure}[h]
\lstset{language=C,xleftmargin=3em,frame=single,framexleftmargin=0em,
basicstyle=\footnotesize\ttfamily,
  keywordstyle=\color{blue}, frame=single,numbers=left,
}%
\begin{lstlisting}
param N;
statement S1a, S1b, S1c;
statement S2a, S2b, S2c;
statement S3; 
statement S4a, S4b, S4c;
statement S5a, S5b, S5c;

S1a := {i: 0<=i<N (%) #CNOT(i, i+1) };
S1b := {i: 0<=i<N (%) #CNOT(i+1, i) };
S1c := {i: 0<=i<N (%) #CNOT(i, i+1) };

S2a := {i: 0<=i<N (%) #CNOT(2*N-i+1, 2*N-i) };
S2b := {i: 0<=i<N (%) #CNOT(2*N-i, 2*N-i+1) };
S2c := {i: 0<=i<N (%) #CNOT(2*N-i+1, 2*N-i) };

S3 := {i: i = N (%) #CNOT(i, i+1) (+) 
  #CNOT(i+1, i) (+) #CNOT(i, i+1) };

S4a := {i: 0<=i<N (%) #CNOT(N-i-1, N-i) };
S4b := {i: 0<=i<N (%) #CNOT(N-i, N-i-1) };
S4c := {i: 0<=i<N (%) #CNOT(N-i-1, N-i) };

S5a := {i: 0<i<=N (%) #CNOT(N+i+1, N+i) };
S5b := {i: 0<i<=N (%) #CNOT(N+i, N+i+1) };
S5c := {i: 0<i<=N (%) #CNOT(N+i+1, N+i) };

codegen { S1a (+) S1b (+) S1c (+) S2a (+) 
  S2b (+)  S2c (+) S3 (+) S4a (+) S4b (+) 
  S4c (+) S5a (+) S5b (+) S5c } with { N=6};
\end{lstlisting}
\caption{\label{lst:axl-pipelined}\axl~ implementation of \bpipelined~ circuit}
\end{figure}